\documentclass[journal,twoside,web]{ieeecolor}
\usepackage{tmi}
\usepackage{cite}
\usepackage{amsmath,amssymb,amsfonts}
\usepackage{algorithmic}
\usepackage{graphicx}
\usepackage{textcomp}
\usepackage{bm}
\usepackage{array}
\usepackage{xurl}

\usepackage{verbatim}
\usepackage[colorlinks,
linkcolor=blue,
anchorcolor=blue,
citecolor=blue]{hyperref}

\makeatletter
\renewcommand{\maketag@@@}[1]{\hbox{\m@th\normalsize\normalfont#1}}%
\makeatother

\def\BibTeX{{\rm B\kern-.05em{\sc i\kern-.025em b}\kern-.08em
		T\kern-.1667em\lower.7ex\hbox{E}\kern-.125emX}}
\begin{document}
	\title{Joint Progressive and Coarse-to-fine Registration of Brain MRI via Deformation Field Integration and Non-Rigid Feature Fusion}
	\author{Jinxin Lv, Zhiwei Wang, Hongkuan Shi, Haobo Zhang, Sheng Wang, Yilang Wang, and Qiang Li, \IEEEmembership{Member, IEEE}
		\thanks{Jinxin Lv, Zhiwei Wang, Hongkuan Shi, Haobo Zhang, Sheng Wang, Yilang Wang, Qiang Li are with Britton Chance Center for Biomedical Photonics, Wuhan National Laboratory for Optoelectronics and with MoE Key Laboratory for Biomedical Photonics, Collaborative Innovation Center for Biomedical Engineering, School of Engineering Sciences, Huazhong University of Science and Technology, Wuhan, Hubei 430074, China. Jinxin Lv and Zhiwei Wang are the co-first authors contributing equally to this work. Qiang Li is the corresponding author. (email:liqiang8@hust.edu.cn)}}
	
	\maketitle
	
\begin{abstract}
Registration of brain MRI images requires to solve a deformation field, which is extremely difficult in aligning intricate brain tissues, e.g., subcortical nuclei, etc. Existing efforts resort to decomposing the target deformation field into intermediate sub-fields with either tiny motions, i.e., progressive registration stage by stage, or lower resolutions, i.e., coarse-to-fine estimation of the full-size deformation field. In this paper, we argue that those efforts are not mutually exclusive, and propose a unified framework for robust brain MRI registration in both progressive and coarse-to-fine manners simultaneously. Specifically, building on a dual-encoder U-Net, the fixed-moving MRI pair is encoded and decoded into multi-scale sub-fields from coarse to fine. Each decoding block contains two proposed novel modules: i) in \emph{Deformation Field Integration (DFI)}, a single integrated deformation sub-field is calculated, warping by which is equivalent to warping progressively by sub-fields from all previous decoding blocks, and ii) in \emph{Non-rigid Feature Fusion (NFF)}, features of the fixed-moving pair are aligned by DFI-integrated deformation field, and then fused to predict a finer sub-field. Leveraging both DFI and NFF, the target deformation field is factorized into multi-scale sub-fields, where the coarser fields alleviate the estimate of a finer one and the finer field learns to make up those misalignments insolvable by previous coarser ones. The extensive and comprehensive experimental results on both private and two public datasets demonstrate a superior registration performance of brain MRI images over progressive registration only and coarse-to-fine estimation only, with an increase by at most 8\% in the average Dice.
\end{abstract}


{\tiny }\section{Introduction}
\label{sec:introduction}
\begin{figure}[!t]
	\centerline{\includegraphics[width=1.0\columnwidth]{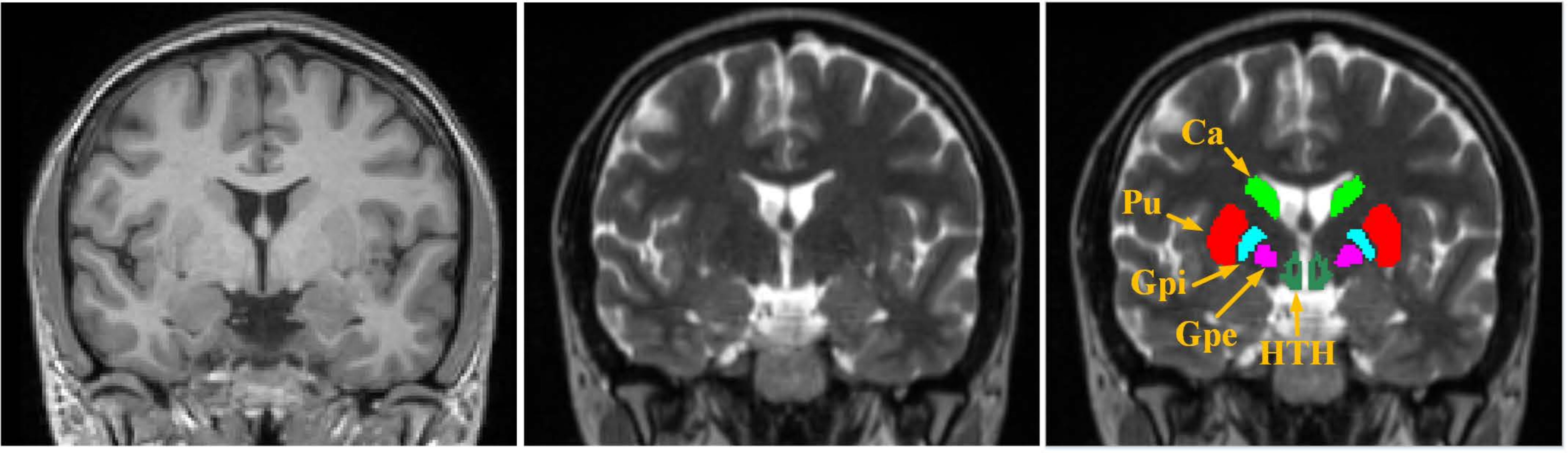}}
	\caption{From left to right: T1w image, T2w image and T2w image overlapped with ground-truth masks of subcortical nuclei.}
	\label{fig_1}
\end{figure}

\IEEEPARstart{A}{ccurate} registration of brain Magnetic Resonance Imaging (MRI) images allows the neurosurgeon to trace brain areas of interest, e.g., subcortical nuclei, brain functional areas, tumors, etc., across different modalities or different patients for the diagnosis of varied neurological diseases. For instance, deep brain stimulation (DBS), as the most developed treatment of Parkinson’s disease, asks to electrically stimulate the subcortical nuclei of STN, GPi and GPe. These nuclei are hard to pre-locate due to blur edges, low contrasts and small scales as evidenced in Fig. \ref{fig_1}. The most common and reliable clinical strategy is to propagate those cross-checked ground-truth masks, which were pre-defined on a MRI atlas with higher magnetic flux density, to the target MRI images via registration, namely atlas-based segmentation \cite{cite1}. Similar procedure is also widely-applied to the diagnosis of Alzheimer's disease (AD) for analyzing the changes in the volume of the hippocampus.

The objective of brain MRI registration is to search a deformation field, which defines a voxel-to-voxel displacement map between the moving image and the fixed image to spatially align them. Traditional registration methods \cite{cite2,cite3,cite4,cite5,cite6,cite7,cite8,cite9,cite10} typically search the deformation field by iteratively optimizing an empirically-formulated energy function defined by the warped and the fixed images. Such iterative optimization process is often time-consuming, and consequently impedes the application of those methods in real-time scenarios, e.g., image-guided intraoperative navigation.


Recently, learning-based methods \cite{cite14,cite15,cite16,cite55} have proliferated and proposed to directly estimate the deformation field in an optimization-free inference phase. These methods typically start from establishing a parameterized mapping function from a fixed-moving image pair to its target deformation field, and then train the parameters based on a collection of unregistered pairs. Once trained, a deformation field can be inferred by a single feed forward in no time. Despite their computational efficiency, if the displacement between the paired images is too large, the deformation field could be too difficult to estimate by once, which often invalidates those learning-based methods accordingly \cite{cite21}. Several works \cite{cite21,cite22} have also observed the same problem and argued that the performance of direct estimation approaches \cite{cite15,cite16} is usually limited on the challenging clinical applications where large displacements occur \cite{cite56}.

To address this, the potential solution is to decompose a target deformation field into several simpler ones, each of which can be easier to estimate. By decomposition, it is allowed to warp a moving image continuously by many times and keep correcting the misalignment during the movement. Thus, the misalignment unsolved by the previous decomposed deformation fields could be tolerated and corrected in the next estimation. Zhao \emph{et al}. \cite{cite21} proposed to decompose the target deformation field into multiple sub-fields in a progressive manner by a set of cascaded CNNs named VTN. Each CNN learns to re-align the warped image by previous CNN to the fixed image. Despite of its high accuracy of registration, it comes at the expense of computational efficiency. Likewise, Hu \emph{et al}.\cite{cite22} proposed to decompose the target deformation field in a coarse-to-fine manner via a Dual-PRNet, where a coarse sub-field is predicted by the shallow layer to help estimate a fine one in the next layer. Such coarse-to-fine decomposition can achieve a promising registration performance with much fewer parameters. However, Dual-PRNet decomposes the deformation field heavily relying on every two adjacent sub-fields, and thus leaves a great room for a further improvement.


In this paper, we take a further step on both works of VTN and Dual-PRNet, and propose a unified framework for robust brain MRI registration in both progressive and coarse-to-fine manners simultaneously. Our motivation is based on an argument that more lightweight CNNs or even CNN layers, i.e., decoding blocks in our implementation, could be competent in the progressive decomposition if the coarse-to-fine estimation is also taken into consideration. That is to say, all the previously predicted coarser fields can be utilized to warp features progressively to help estimate the next finer field. 

\begin{figure*}[!t]
	\centerline{\includegraphics[width=0.96\textwidth]{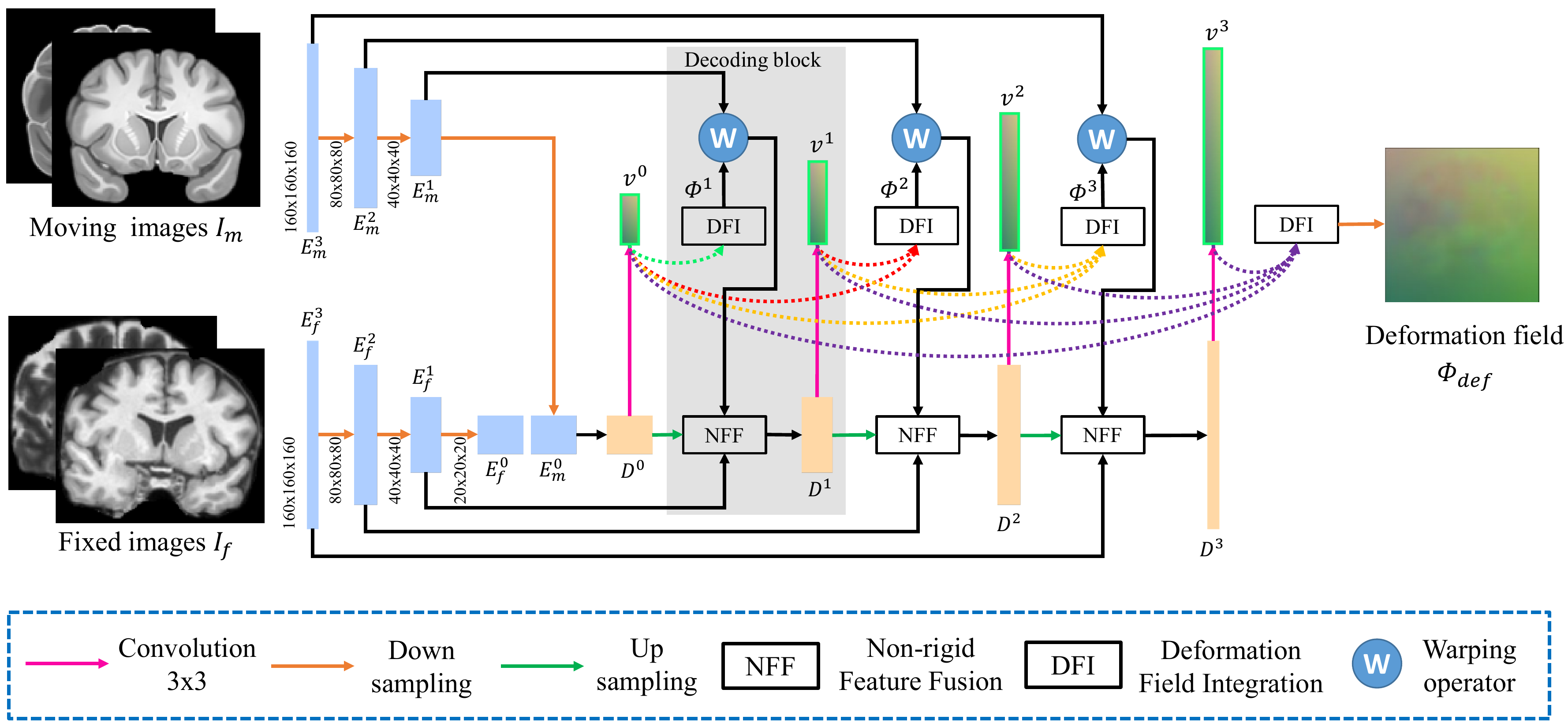}}
	\caption{The structure of the unified deformable registration framework, which consists of a dual-encoder U-Net backbone, and a set of DFI and NFF modules to decompose the target deformation field into multi-scale sub-fields. $I_m$ and $I_f$ represent the concatenated moving images and concatenated fixed images respectively, i.e., $I_m=cat(I_m^{T1w}, I_m^{T2w})$, $I_f=cat(I_f^{T1w}, I_f^{T2w})$. $\Phi^{1}, \Phi^{2}, \Phi^{3}$ represent the DFI-integrated deformation fields to warp the moving image's feature. $\Phi_{def}$ is the final predicted deformable deformation field. The warping operator is based on Spatial Transformer Networks.}
	\label{fig_2}
\end{figure*}

To this end, we employ a dual-encoder U-Net \cite{cite22} to first encode hierarchical feature maps of the moving and fixed images separately. As shown in Fig. \ref{fig_2}, the two feature maps at the last encoding layer are then fused and decoded into multi-scale sub-fields sequentially. Inspired by \cite{cite16}, these sub-fields are defined as velocity fields instead of displacement maps to force a diffeomorphic transformation. In each decoding block, DFI first up-scales all previous coarser velocity sub-field(s) to the same size, and then fuse them via a spatial attention to obtain a single velocity field, warping by which is equivalent to warping progressively by its constituent sub-fields. The final integrated deformation field is obtained by integrating the fused velocity field over unit time \cite{cite16}. Then NFF utilizes the integrated field to “progressively” deform the skip-connected encoding feature map of the moving image, and fuses the aligned moving-fixed encoding feature maps and the previous decoding feature map to obtain the next decoding feature map. Finally, the NFF-fused feature map is convolved and transformed to a finer velocity sub-field which learns to solve those misalignments remaining by all previous coarser velocity sub-fields.

To summarize, our key contributions are as follows:
\begin{itemize}
	
	\item \emph{We propose a unified framework for a robust registration of brain MRI by decomposing the single difficult deformation field in both progressive and coarse-to-fine manners jointly.}
	
	\item \emph{We propose two novel modules named Deformation Field Integration (DFI) and Non-rigid Feature Fusion (NFF) to decompose the full-size deformation field in all decoding blocks instead of just two adjacent layers or multiple heavy-weight CNN models.}
	
	\item \emph{Extensive and comprehensive experimental results on both public and private datasets demonstrate the superior registration performance of our unified framework and the effectiveness of our proposed two modules, i.e., DFI and NFF. The source code of our implementation is available\footnote{\url{https://github.com/OldDriverJinx/Progressvie-and-Coarse-to-fine-Registration-Network}}.}
\end{itemize}

\section{Related work}

\subsection{Registration Problem}\label{sec_3a}

The registration problem is to solve a spatial transformation or mapping between two (or more) medical images so that the corresponding points in different images can be uniformly aligned \cite{cite58}. Depending on the clinical demand, the registration problem can be categorized as three classes, i.e., cross-modal registration, cross-patient registration and cross-time registration. Cross-modal registration aims at aligning the same tissue structure in images from different modalities, e.g., CT, MRI, etc., and it enables multi-modal fusion or navigation in Image-Guided Surgery (IGS). For instance, registration of preoperative MRI and intraoperative ultrasound \cite{cite59,cite60} provides the surgeon with more accurate information about the shape and location of the target lesions. Cross-patient registration, which this paper belongs to, aims at aligning the small and intricate structures with different shapes and/or sizes among patients. It is widely-used in brain atlas analysis \cite{cite34,cite38}. Cross-time registration asks to align a set of sequential images scanned in different time. It can help doctors to eliminate large motions, e.g., heart movement \cite{cite61}, pulmonary respiratory motion estimation \cite{cite62}, or to observe the changes of Regions of Interest (RoIs).

In general, an accurate deformation field is the key to solve the above three registration problems. Existing methods mainly resort to optimization-based approaches (i.e., traditional registration methods), or learning-based approaches to estimate the deformation field.


\subsection{Traditional Registration Methods}\label{sec_3b}

Traditional registration methods typically define an energy function to measure the similarity between two images, and iteratively search a parametrized deformation field to make the energy function optimal with input of the warped and the fixed images. These methods differ from each other in terms of mathematically modeling the deformation field in different ways, e.g., elastic-type models \cite{cite50}, free-form deformation model based on the cubic B-spline \cite{cite51} (integrated in Elastix \cite{cite10}), statistical parametric mapping \cite{cite52}, and Demons \cite{cite53}. To make the deformation field smoother and invertible, the diffeomorphic transformations have attracted numerous studies in the past decade \cite{cite45}, e.g., Large Diffeomorphic Distance (LDDMM) \cite{cite46,cite47}, Symmetric Normalization (SyN) \cite{cite9} (integrated in ANTs \cite{cite42}) and diffeomorphic Demons \cite{cite48}. Despite their success, they share a common flaw of high computing cost caused by the iterative optimization for medical images, not mention to high-dimensional 3D brain MRI data. It might cost minutes (by Elastix) or even hours (by ANTs) \cite{cite16} for the registration of two MRI images with the size of 160×160×160.


\subsection{Learning-based Registration Methods}\label{sec_3c}

The learning-based methods typically parameterize a mapping relation (e.g., CNN) from the fixed-moving image pair to the target deformation field, and learn this parameterized mapping via a number of unregistered training image pairs. Compared with the traditional optimization-based approaches, the time cost of learning-based method is mainly in the training stage. Once trained, the inference phase is optimization-free, and thus extremely fast (around 1 second per 3D image pair with the size of 160×160×160). 

Ground-truth deformation fields for supervised training are either obtained by traditional approaches, or synthesized by manual deformation transformations. For instance, Fan \emph{et al.} \cite{cite11} used the deformation field obtained by LCC-Demons \cite{cite12} and SyN \cite{cite9} to train a CNN. Zhu \emph{et al.}\cite{cite13} first randomly simulated deformation field including rotation, scale, translation and elastic deformation, and then applied it to the actual image to obtain the synthetic moving image, and finally used the generated deformation field to supervise the training of the network. However, those pseudo ground-truth deformation fields are mostly, if not all, unable to reveal the real distribution of misalignments, and thus lead to an unpromising registration performance of the supervised methods.

To address this problem, the unsupervised registration methods were proposed to rely on the image-wise similarities instead of the synthesized unreal deformation fields for training. Fan \emph{et al.}\cite{cite14} introduced a GAN-induced measurement as the similarity between the moved and fixed images, and forced a registration network to maximize it during the learning. Balakrishnan \emph{et al.}\cite{cite15} developed a U-Net  \cite{cite27} liked framework named VoxelMorph and trained it using both volume-level similarity loss, i.e., normalized cross-correlation (NCC) or voxel-level similarity loss, i.e., mean squared error (MSE). Dalca \emph{et al.}\cite{cite16} extended the VoxelMorph, and used a probabilistic generative model named VoxelMorph-diff to generate a diffeomorphic deformation field, well preserving the topology of the moving image.

Despite their success, the major problem of the unsupervised approaches is the sub-optimal registration performance for those important internal areas of interest (e.g., brain nuclei). To this end, the weakly supervised methods were proposed to also introduce the spatial consistency between segmented areas of interest  \cite{cite17,cite18} and/or between anatomical landmarks \cite{cite19,cite20} as the training guidance. By benefits of those additional constraints, existing methods further push forward the registration performance. For instance, the authors of VoxelMorph \cite{cite15,cite16} have demonstrated that additionally optimizing the segmentation loss can improve the registration performance in RoIs. However, a few efforts touch the core daunting challenge that direct estimation of the full-size deformation field is difficult on its own and hardly eased no matter how many supervisions are introduced.

In view of this, several methods attempted to alleviate the difficulty by decomposing the single deformation field into several sub-fields in either progressive manner or coarse-to-fine manner.

\textbf{Progressive registration} incrementally moves the moving image with a tiny motion every time, warping it onto the fixed image gradually. Volume Tweening Network (VTN) proposed by Zhao \emph{et al.}\cite{cite21} decomposed the large deformation into a series of deformation sub-fields by several cascaded CNNs. After three times of progressively non-rigid warping, the moved image finally gets close to the fixed image. Zhao \emph{et al.}\cite{cite28} also proposed to enlarge the number of cascades to push the limit of VTN, and demonstrated that, as the number of cascades increases, the performance improves accordingly. Eventually, 10-cascade VTN achieved the highest accuracy of registration comparing with existing learning-based methods for abdominal CT images and brain MRI images. However, the side effect of increasing the number of cascades is the massive parameters involved inevitably, which makes such decomposition over-parametrized.

\textbf{Coarse-to-fine estimation} starts from predicting a lower-resolution (coarser) deformation field where neighboring voxels are grouped and share the same displacement vector, then estimates a finer field where fewer voxels are grouped, and finally yields the deformation field with the same resolution of the moving image. Hu \emph{et al.}\cite{cite22} developed a dual-stream pyramid registration network (Dual-PRNet) for a coarse-to-fine estimation, which uses the previous coarse field to align features of the moving and fixed images so as to help obtain the next fine field. Comparing with both direct estimation method, i.e., VoxelMorph \cite{cite15} , and progressive registration method, i.e., 10-cascade VTN \cite{cite28}, Dual-PRNet can achieve a medium registration performance but with much fewer parameters.




In summary, the progressive decomposition in VTN relies on multiple large CNN models, yielding a superior registration performance but sacrificing the computational efficiency. The coarse-to-fine decomposition in Dual-PRNet relies on a single CNN’s multi-scale feature maps, only requiring much fewer parameters for a promising registration performance. Their strengths can be compatibly combined since coarse-to-fine decomposition can significantly lower the high computational demand of VTN, and progressive warping by those decomposed sub-fields can further improve the registration performance of Dual-PRNet.

Inspired by this, we in this paper build a unified framework for robust brain MRI registration in both progressive and coarse-to-fine manners simultaneously. Thanks to the joint progressive and coarse-to-fine decomposition, our method can achieve a superior registration performance while only increases a negligible computational cost.

\begin{figure}[!t]
	\centerline{\includegraphics[width=1.0\columnwidth]{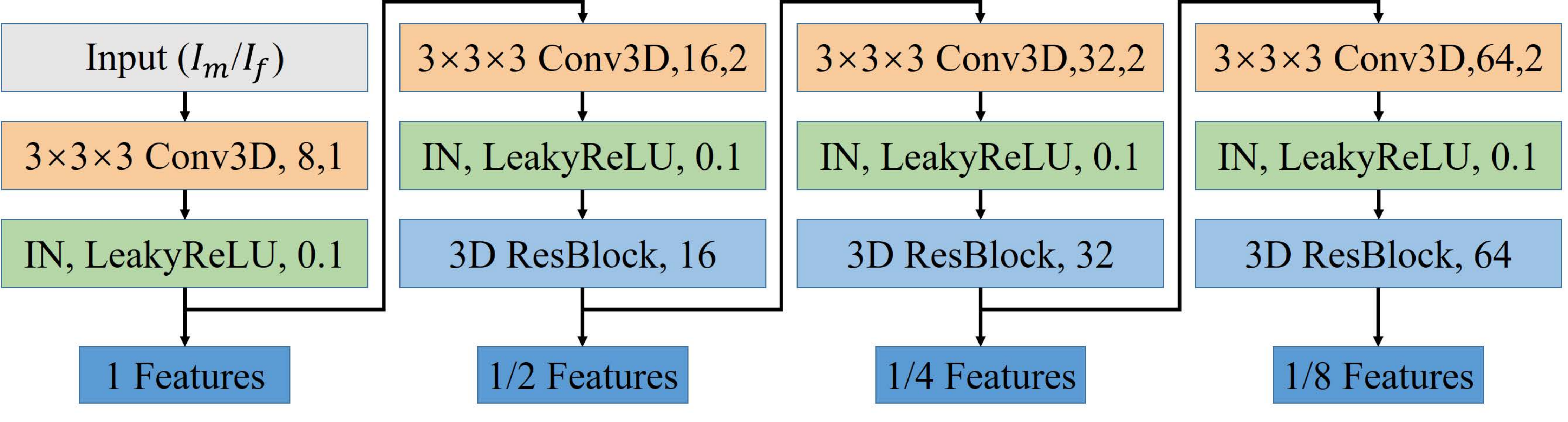}}
	\caption{The architecture of each encoder of our dual-encoder. \emph{3x3x3 Conv3D,8,1} indicates a 3D convolutional layer using 8 kernels with the size of 3×3×3 and the stride of 1. \emph{IN, LeakyReLU, 0.1}: an Instance Normalization layer with a LeakyReLU activation function with the slope of 0.1. \emph{3D ResBlock, 16}: 3D ResBlock with 16 output channels. \emph{1/n Features}: a side-output feature map down-sampled by n times.}
	\label{fig_3}
\end{figure}

\section{Method}
Fig. \ref{fig_2} illustrates our unified framework, each decoding block of which consists of two major modules, i.e., Deformation Field Integration (DFI) and Non-rigid Feature Fusion (NFF). In the following, we first introduce the backbone of our framework, i.e., the dual-encoder U-Net (Sec. \ref{sec_2a}), then detail both DFI and NFF (Sec. \ref{sec_2b}\&\ref{sec_2c}), and at last give implementation and training details (Sec. \ref{sec_2d}). 

\subsection{Unified Framework for Deformable Registration}\label{sec_2a}

\textbf{Input:} As shown in Fig. \ref{fig_2}, the atlas and target brain MRIs are considered as moving and fixed MRIs respectively, that is, our goal is to predict a deformation field which can warp the moving MRI onto the fixed MRI. Each brain MRI data consists of two modalities, i.e., T1w and T2w images which were already spatially aligned beforehand. We employ the dual-encoder U-Net  \cite{cite22} as our backbone, and thus directly concatenate the two (T1w and T2w) images to fit the input of the backbone. The concatenated moving images and concatenated fixed images are denoted as $I_m$ and $I_f$ respectively, i.e., $I_m=cat(I_m^{T1w}, I_m^{T2w})$, $I_f=cat(I_f^{T1w}, I_f^{T2w})$, $cat(\cdot)$ means concatenation operation.

\textbf{Dual-encoder:} Given an unregistered pair of $I_m$ and $I_f$, the dual-encoder U-Net, as its name implies, utilizes two separate encoders to extract hierarchical features from $I_m$ and $I_f$ respectively. Fig. \ref{fig_3} illustrates one of two identical branches of the dual-encoder, which contains four encoding blocks. Each encoding block is represented by the three boxes colored in orange, green, blue, respectively (the leftmost two boxes colored in orange and green represent the first encoding block). Specifically, the first encoding block consists of a convolution layer, Instance Normalization (IN) layer \cite{cite57} with a LeakyReLU activation function, yielding the first side-output feature map with the same size of input image. Each of the last three encoding blocks consists of a convolution layer with the stride of 2, IN layer, and a ResBlock \cite{cite23} to extract a side-output feature map with the size down-scaled by twice.

The four encoding blocks totally produce four side-output feature maps as shown in Fig. \ref{fig_3}. Thus, the dual-encoder extracts two sets of encoding feature maps denoted as \{$E_{m}^{L-1}, \cdots, E_{m}^{0}$\},\{$E_{f}^{L-1}, \cdots, E_{f}^{0}$\} for $I_m$ and $I_f$ respectively, where $L=4$, $E^l$ represents the feature map from $l$-th encoding layer, $L-1$ and $0$ mean the shallowest and the deepest layer respectively, that is, the volume of $E^{l}$ is $2^3$ times larger than that of $E^{l-1}$, and $E^{L-1}$ has the same size as $I$.


We would highlight the advantage of usage of dual-encoder comparing to the single-encoder employed in both VoxelMorph\cite{cite15} and VTN \cite{cite21}. As can be seen in Fig. \ref{fig_2}, the discrepancy of image style between $I_m$ and $I_f$ is non-negligible, while the single-encoder concatenates $I_m$ with $I_f$ at the very first layer, which inevitably entangles both spatial misalignment and style discrepancy at those shallow layers, and embarrasses the decoder consequently. The decoder following a single-encoder not only has to reveal a deformation field from the encoded spatial misalignment, but also should keep itself from being disturbed by the entangled style discrepancy. By contrast, the dual-encoder encodes appearances of $I_m$ and $I_f$ separately without image style entangled, yielding encoding feature maps more friendly for the estimation of deformation map \cite{cite25,cite26}.

\textbf{Single-decoder:} We concatenate two encoding feature maps from the last layer, i.e., $cat(E^0_m,E^0_f)$, for decoding. Similar to the encoding phase, a set of decoding feature maps are obtained \{$D^{0}, \cdots, D^{L-1}$\}, where $D^{l}$ has the same size with $E^l$ and is calculated by a decoding block except for $D^0$. $D^0$ is obtained by directly convolving the concatenated encoding feature maps as shown in Fig. \ref{fig_2}.

Each decoding feature map first goes through a convolution layer to produce a corresponding velocity sub-field which gives a set of multi-scale velocity sub-fields, denoted as \{$v^{0}, \cdots, v^{L-1}$\}, where $v^l=conv(D^l)$, and the $conv(\cdot)$ represents the convolution operation which utilizes three convolutional kernels with size of $3\times3\times3$ to transform $D^l$ to a 3-channel $v^{l}$, velocity vector map for x, y, z directions. Technically, the decoder also can directly generate the deformation sub-fields. We instead predict these intermediate velocity sub-fields with the purpose of forcing the smoothness and diffeomorphism in the converted deformation field by DFI.


\begin{figure}[!t]
	\centerline{\includegraphics[width=1.0\columnwidth]{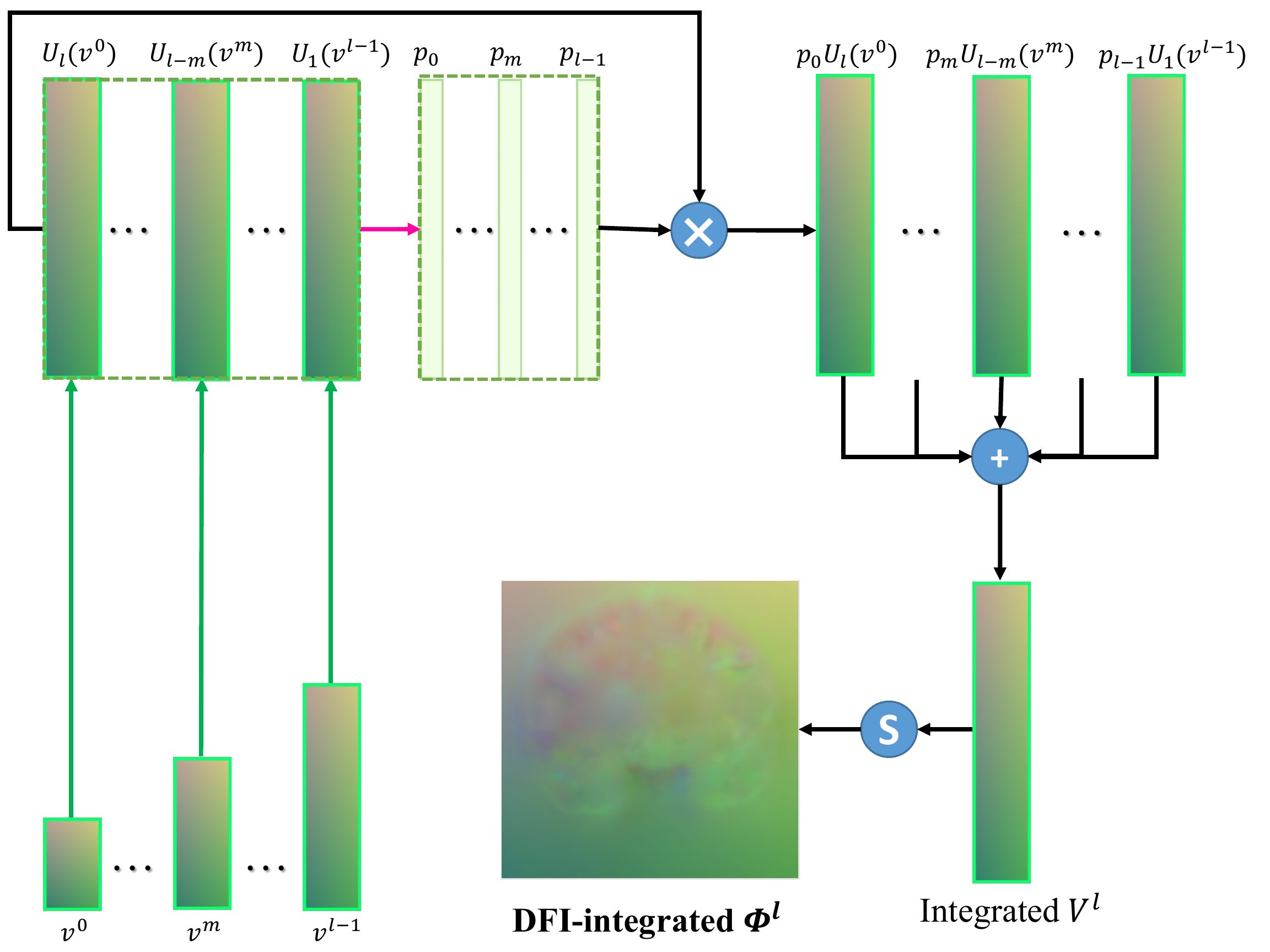}}
	\caption{The structure of DFI, which integrates sub-fields from all previous decoding blocks. The pink arrow represents the convolution operation with $3\times3\times3$ kernel size. The green arrow represents the up-sampling operation. The cross sign denotes element-wise multiply operation. The symbol 'S' represents a recurrent warping scheme in Eq. \eqref{eq_2}. }
	\label{fig_4}
\end{figure}

Combining both dual-encoder and single-decoder, our method not only extracts the registration-friendly encoding features of the moving and fixed images, but also decodes sub-fields based on those deeply fused features to prevent overfitting to a specific image pattern (e.g., the same fixed image). In comparison, Dual-PRNet separately extracts features of moving and fixed images from the beginning to end via a dual-encoder and a dual-decoder, which probably hurts the generalization performance since separate feature learning could make the model overfit to a specific intensity pattern, especially when the fixed image keeps unchanged during the training. Please refer to Sec. \ref{sec_4e} for the discussion in detail.



In the following subsections, we explain how to calculate $D^l$ in the decoding block using both DFI and NFF.

\subsection{Deformation Field Integration Module}\label{sec_2b}

As shown in Fig. \ref{fig_4}, DFI in $l$-th decoding block takes velocity sub-fields from all previous blocks as inputs, and then fuses them into a single velocity field \{$v^{0}, \cdots, v^{l-1}$\}$\rightarrow V^{l}$ via a spatial attention mechanism. Lastly, the fused $V^{l}$ is converted to the integrated deformation field $\Phi^l$. Fig. \ref{fig_4} illustrates the concrete procedure of DFI.

\begin{figure}[!t]
	\centerline{\includegraphics[width=0.9\columnwidth]{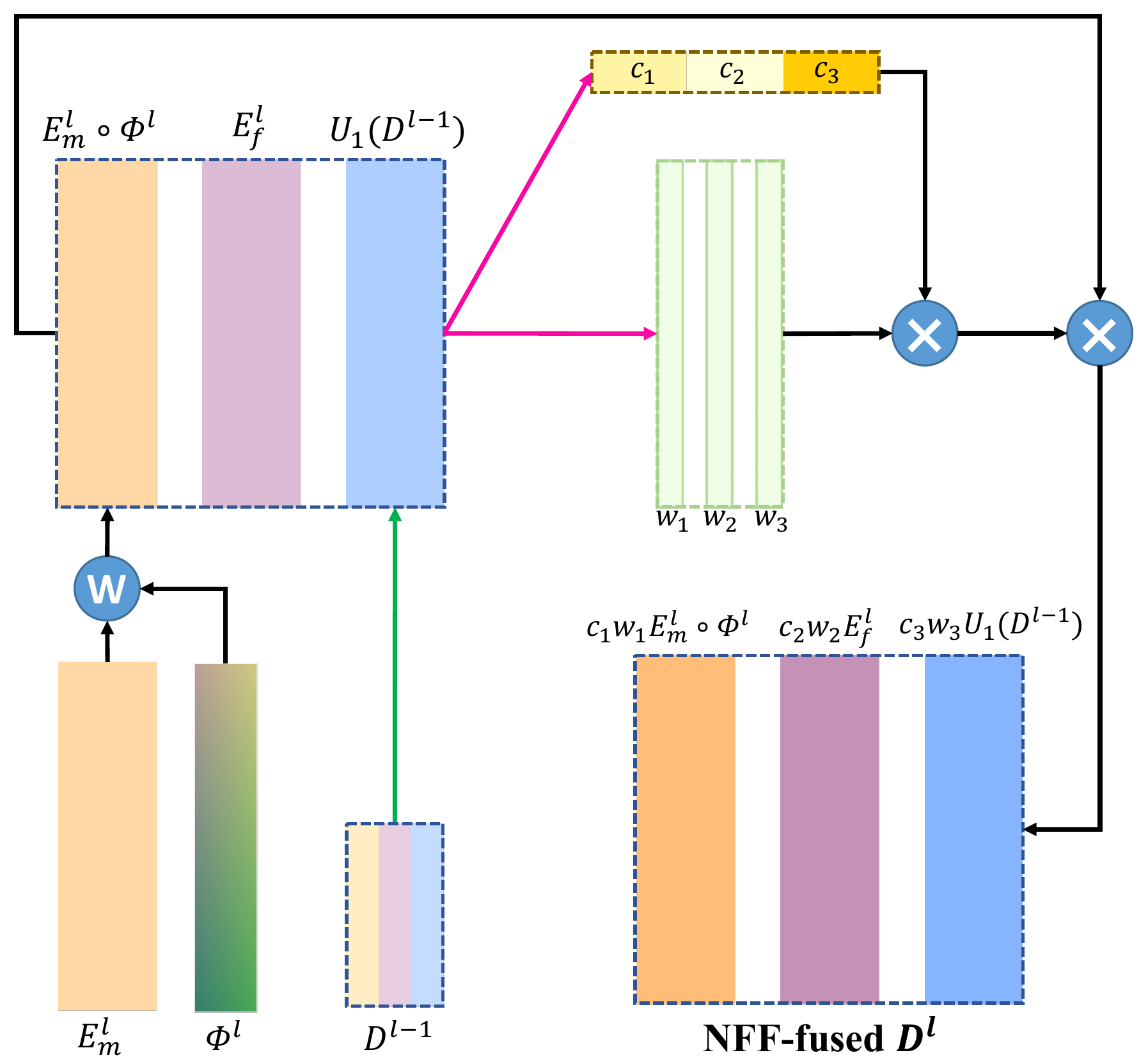}}
	\caption{The structure of NFF, which dynamically fuses features from three sources with channel and spatial attention.}
	\label{fig_5}
\end{figure}

\begin{figure*}[t]
	\centerline{\includegraphics[width=1.0\textwidth]{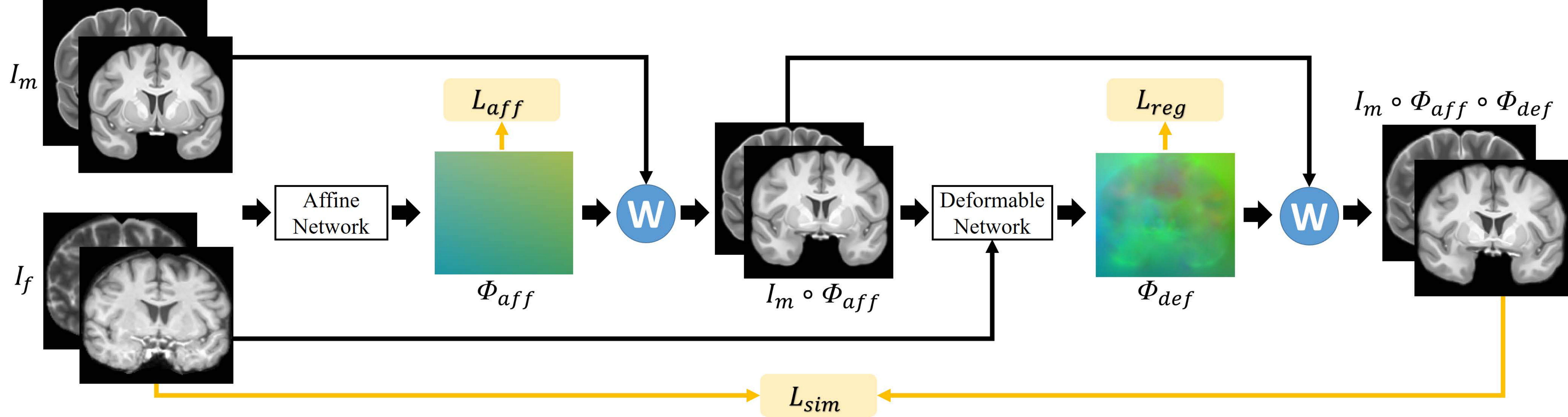}}
	\caption{The implementation of our cascaded method consisting of affine and deformable parts, where the deformable part is our proposed unified framework.}
	\label{fig_6}
\end{figure*}


Note that, the confidence of the decoding block varies under different scales and different locations. For instance, large displacement vectors are easier to estimate at a lower resolution due to large receptive fields, and fine ones, conversely, are suitable to estimate at a higher resolution. Therefore, $l$ up-scaled velocity sub-fields are concatenated and convolved into $l$ weighting maps \{$p_0, …, p_{l-1}$\} to re-weight the contribution of each velocity vector in the corresponding velocity sub-field, i.e., $v'^m = p_m U_{l-m}(v^m)$. Those weighting maps are expected to filter out those untrustworthy predictions, yielding more accurate and robust integrated deformation field. $V^l$ is therefore calculated as:

\begin{equation}\label{eq_1}
	V^{l}=\sum_{m=0}^{l}v'^m
\end{equation}


$\Phi^l$ is obtained by integrating $V^l$ over [0,1] inspired by VoxelMorph-diff \cite{cite16}. Specifically, we first define a very small time step $1/2^{t}$, where $t=7$ in our implementation, and employ a recurrent warping scheme:

\begin{equation}\label{eq_2}
	\phi^l_{\left( 1/2^{t-1} \right)} = \phi^l_{\left( 1/2^t \right)} \circ \phi^l_{\left( 1/2^t \right)},
\end{equation}
where $\circ$ is the warping operation defined as $I \circ \phi = I(\bm{p}+\phi(\bm{p}))$, $\bm{p}$ is a map of spatial locations.

We let $\phi^l_{\left( 1/2^{t} \right)} = V^l/2^t $ as the start of the recurrent warping in Eq. (\ref{eq_2}), and eventually obtain the final integrated deformation field $\Phi^l = \phi^l_{\left(1\right)} = \phi^l_{\left( 1/2 \right)} \circ \phi^l_{\left( 1/2 \right)} $. Warping by $\Phi^l$ is equivalent to warping ``progressively" by the constituents of $V^l$, which are all previous sub-fields \{$v^{0}, \cdots, v^{l-1}$\}.

\subsection{Non-rigid Feature Fusion module}\label{sec_2c}


The $l$-th decoding block in the dual-encoder U-Net fuses feature maps from three sources, i.e., from the two counterpart encoders ($E_m^l$ and $E_f^l$) and from the last decoder ($D^{l-1}$). By comparing $E_m^l$ and $E_f^l$, a deformation field can be inferred from the underlying hints of misalignments between fixed and moving images. This forms the basic idea of VoxelMorph \cite{cite15}. However, we believe that the inference can be eased by first progressively aligning $E_m^l$ and $E_f^l$ instead of directly comparing them. That is, those misalignments are largely corrected by previous sub-fields beforehand, and the current decoding block can thus focus on predicting a finer sub-field for unsolved misalignments.

Specifically, as shown in Fig. \ref{fig_5}, the DFI-integrated deformation field $\Phi^{l}$ is first used to non-rigidly transform $E_m^l$, i.e., $ E_m^l \circ\Phi^l $. Second, decoding feature map from the last block $D^{l-1}$ is up-scaled using a transposed convolution\cite{cite36} with the kernel size and the stride equal to 4 and 2 respectively, i.e., $U_1(D^{l-1})$. Third, NFF fuses the three feature maps to a new decoding feature map \{$ E_m^l\circ \Phi^l , E_f^l, U_1(D^{l-1})$\} $\rightarrow$ $D^l$ by both channel and spatial attentions inspired by \cite{cite29}.

Fig. \ref{fig_5} illustrates two separate branches utilized by NFF to calculate channel-wise and spatial-wise attentions respectively. One branch yields a soft-maxed attention vector $c$ whose number of channels is the sum of the inputs'. Another branch yields three soft-maxed attention maps \{$w_1,w_2,w_3$\}. We divide $c$ into three parts $\left\{c_1,c_2,c_3\right\}$ along channel with guarantee that \{$  E_m^l \circ \Phi^l , E_f^l, U_1(D^{l-1})$\} and \{$c_1,c_2,c_3$\} are orderly matched with each other in terms of the channel. \{$w_1,w_2,w_3$\} and \{$c_1,c_2,c_3$\} are then broadcasted and multiplied together with \{$ E_m^l \circ \Phi^l , E_f^l, U_1(D^{l-1})$\} to obtain $D^l$ as:

\begin{equation}\label{eq_3}
	D^{l}=cat( c_1 w_{1} E_{m}^{l}\circ \Phi^{l}  , c_2 w_{2} E_{f}^{l}, c_3 w_{3}U_{1}(D^{l-1})).
\end{equation} 

By doing so, the soft-maxed attention vectors and maps can normalize feature scales from different sources and highlight those more contributing ones at different locations and channels.

We would highlight the differences between our method and Dual-PRNet \cite{cite22} which also proposed to non-rigidly deform features for estimation. First, in Dual-PRNet features of the moving and fixed images are learned separately and interact only for prediction of deformation sub-fields, while our method utilizes NFF to fuse them as an auxiliary information to assist the prediction. Second, each sub-field in \cite{cite22} is predicted based on deformed features by the single previous sub-field, that is, the decomposition heavily relies on the two adjacent sub-fields. Our method instead utilizes DFI-calculated integrated $\Phi^{l}$ to warp moving image features, which makes the decomposition job-shared by all $l$ decoding blocks.

\subsection{Implementation and Training Details}\label{sec_2d}

In the implementation as shown in Fig. \ref{fig_6}, we follow \cite{cite21} to cascade an additional light-weight affine registration network on the top of the unified framework. This affine registration network is to align the moving image with an affine transformation for narrowing the spatial misalignment. Both affine and deformable networks can be trained in an end-to-end fashion. 

$I_m$ and $I_f$ are first concatenated, and then the affine registration network sequentially utilizes 5 ResBlocks and a fully connection layer to predict 12 parameters, i.e., a 3×3 matrix $A^*$ and a 3×1 vector $t$. $A^*$ and $t$ defines an affine displacement map formulated as $\Phi_{aff} = A^*\bm{p}+t$, where $\bm{p}$ represents a map of  spatial locations. We then use $\Phi_{aff}$ to warp the $ I_m $ via spatial transform network (STN)\cite{cite24}, which is differentiable and allows gradients to propagate. 

The affinely transformed moving image $ I_m \circ \Phi_{aff} $, as well as $I_f$ are fed into the following unified framework to obtain a set of velocity sub-fields \{$v^0,…,v^{L-1}$\}, which are integrated into the final deformation field $\Phi^{L}$ as presented in Sec. \ref{sec_2b}. Here, we attribute $\Phi^{L}$ as $\Phi_{def}$ for convenience of understanding. Therefore, the moved image can be written as $ I_m \circ \Phi_{aff} \circ \Phi_{def} $.

To train the cascaded affine and deformable network parts, three unsupervised losses are employed as shown in Fig. \ref{fig_6}:



\subsubsection{$L_{aff}$ for constraining $\Phi_{aff}$}
Since the affine registration network aims at roughly aligning by a rotation, translation and only a small scaling, without overly shearing, the transformation matrix $A^*+E$ thus approximates an orthogonal matrix. To force $A^*+E$ to be orthogonal and keep the same chirality of fixed and moving images in the model learning, we use the affine transformation loss proposed in \cite{cite21}:
\begin{equation}\label{eq_4}
	L_{aff}=(-6+\sum_{i=1}^{3}\left(\lambda_{i}^{2}+\lambda_{i}^{-2}\right))+(\operatorname{det}(A^*+E)-1)^{2}.
\end{equation}

The first half of Eq. \eqref{eq_4} is an orthogonality loss where $\lambda_{1,2,3}$ the singular values of $A^*+E$. The behind idea of this loss is that a matrix is orthogonal if and only if all singular values are 1. Hence if $A^*+E$ is completely orthogonal, the orthogonality loss should be 0, otherwise this term can be greater than 0.

The second half of Eq. \eqref{eq_4} is a determinant loss which avoids reflection transform during the learning since the chirality of fixed and moving images should be the same. When the chirality of the two images is the same, the determinant of $A^*+E$ should satisfy $det(A^*+E)>0$. Together with the orthogonality requirement, the determinant of $A^*+E$ should be closed to 1. Therefore, optimizing $L_{aff}$ forces the invertibility of the predicted affine transformation matrix. Even if the loss is not perfectly optimized, there is little chance that $det(A^*+E)$ happens to be zero, i.e., the predicted matrix is not an affine transformation matrix, since the network is randomly initialized. Experimental details are discussed in Sec. \ref{sec_4d}.


\subsubsection{$L_{reg}$ for smoothness of $\Phi_{def}$}
Inspired by \cite{cite15}, we construct $L_{reg}$ as Eq. (\ref{eq_5}) to ensure the continuity of the deformation field based on its spatial gradients:

\begin{equation}\label{eq_5}
	L_{reg}=\frac{1}{N}\sum(\nabla_{x}\Phi_{def})^{2}+(\nabla_{y}\Phi_{def})^{2}+(\nabla_{z}\Phi_{def})^{2},
\end{equation}

where $\nabla_{x},\nabla_{y},\nabla_{z}$ represents the difference map in $x,y,z$ direction, and $N$ is the total number of voxels.

\subsubsection{$L_{sim}$ for maximizing similarity between the moved and fixed image}
We use Normalization Local Correlation Coefficient (NLCC) as the similarity metric, which is based on Pearson's Correlation Coefficient:
\begin{equation}\label{eq_6}
	\rho(X,Y)=\frac{\sum\left(X-\bar{X}\right)\left(Y-\bar{Y}\right)}{\sqrt{\sum\left(X-\bar{X}\right)^{2}} \sqrt{\sum\left(Y-\bar{Y}\right)^{2}}},
\end{equation}
where $X$ and $Y$ are two variables, e.g., images.

Instead of directly calculating $\rho(I_m \circ \Phi_{aff} \circ \Phi_{def}, I_f)$, we first extract multiply patches from $I_m \circ \Phi_{aff} \circ \Phi_{def}$ and $I_f$ separately using a $8\times8\times8$ sliding window with 5 voxels overlapped, and then average $-(\rho(X,Y))^2$ for those patches as $L_{sim}$:
\begin{equation}\label{eq_7}
	\begin{aligned}
		L_{sim}=-\frac{1}{N}\sum(&\rho( I_m^{T1w}\circ\Phi_{aff}\circ\Phi_{def} , I_f^{T1w},i)^2+\\
		&\rho(I_m^{T2w}\circ \Phi_{aff} \circ\Phi_{def} , I_f^{T2w},i)^2 ),
	\end{aligned}
\end{equation}
where $\rho(X,Y,i)$ is the $\rho$ coefficient value for the $i$-th patches of $X$ and $Y$.

In addition, we also employ a weakly-supervised loss namely Dice loss for better aligning those intricate brain RoIs:
a


\subsubsection{$L_{seg}$ for weakly-supervised guidance}
The Dice\cite{cite31} coefficient is a commonly adapted metric in the field of image segmentation, and is defined as:

\begin{equation}\label{eq_8}
	Dice(X,Y)=\frac{2|X\cap Y|}{|X|+|Y|},
\end{equation} 
where $X$ and $Y$ represent two segmentation maps (i.e., binary masks), $|X|$ and $|Y|$ represents the number of elements equal to 1 in $X$ and $Y$ respectively.

For the situation of multiple anatomical structures, we compute the Dice loss for each structure, and then average all losses together to get the final Dice loss. Specifically, let $S_f$ = \{$S_f^1, …, S_f^C$\} indicate ground-truth segmentations of a fixed image, and $S_m$ = \{$S_m^1, …, S_m^C$\} indicate ground-truth segmentations of a moving image, where $C$ is the number of anatomical structures. The segmentation loss is calculated as:

\begin{equation}\label{eq_9}
	L_{seg}= -\frac{1}{C} \sum_{c=1}^{C} Dice(S^{c}_{f}, S^{c}_{m}\circ \Phi_{aff} \circ\Phi_{def})
\end{equation} 






The final loss for training can be written as:

\begin{equation}\label{eq_10}
	L_{total}=\alpha_{1} L_{aff}+\alpha_{2} L_{reg}+\alpha_{3} L_{sim}+\alpha_{4} L_{seg},
\end{equation} 
The coefficients of $\alpha_{1}, \alpha_{2}, \alpha_{3}$ in Eq. (\ref{eq_10}) are set to 0.1, 1, 1, which were inherited from the experimental settings adopted in \cite{cite21,cite30}, and $\alpha_{4}$ is set to 2 based on experiments on the IXI dataset (see Sec. \ref{sec_4b}). LeakyReLU\cite{cite37} is used with slope of 0.1 as the activation function in each convolutional layer. The computational environment includes a software toolkit of TensorFlow \cite{cite32}, a GPU resource of NVIDIA RTX 3090, a CPU resource of Intel Xeon Gold 5220R. 



\section{Dataset and Evaluation Metrics}

\subsection{Dataset}\label{sec_3a}
This study includes a private dataset and three public datasets, i.e., CIT168 atlas\cite{cite34}, IXI\footnote{\url{http://brain-development.org/ixi-dataset}} and LPBA\cite{cite38}.

\subsubsection{CIT168 atlas}
The CIT168 atlas is a high-resolution probabilistic in-vivo anatomical atlas of subcortical nuclei. It contains two high-resolution templates with and without skull removed, each of which includes both T1w and T2w images. Masks of 16 subcortical nuclei are provided for both templates.

\subsubsection{IXI Dataset}
The IXI dataset contains 576 registered T1w and T2w MR image pairs, including 184 pairs from Hammersmith Hospital using a Philips 3T MRI system, 319 pairs from Guy’s Hospital using a Philips 1.5T MRI system, and 73 pairs from Institute of Psychiatry using a GE 1.5T MRI system. Skull is removed in all data by FreeSurfer\cite{cite33}. The ground-truth masks of 16 subcortical nuclei are obtained by first automatic delineation using citatlaskit\footnote{\url{https://github.com/jmtyszka/citatlaskit}}, and then manual correction by two local expert neurosurgeons.

\subsubsection{Private Dataset}
The private dataset contains 100 registered T1w and T2w MRI image pairs, which are collected using a United-Imaging 3T MRI system. Same procedure as that in the IXI dataset is performed on the private dataset to obtain the ground-truth masks.

\subsubsection{LPBA dataset}
The LPBA (LONI Probabilistic Brain Atlas) dataset contains 40 T1w images collected from the North Shore-Long Island Jewish Health
System and corresponding ground-truth masks of 56 anatomical structures. These 56 regions include 50 cortical structures, 4 subcortical areas, the brainstem, and the cerebellum. The detail definitions of these 56 regions can refer to \cite{cite38}. All labels were  delineated by trained raters following protocols of LPBA. Skull is also removed in all data by FreeSurfer.

All data are pre-processed to crop useless black areas around the target head and to resample into the size of 128×128×128 voxels for LPBA and 160×160×160 for the rest datasets for compromising the insufficient GPU memory. Intensities are bias-corrected by ANTs\cite{cite42} and linearly normalized to [0,1] by min-max normalization.

\subsection{Evaluation Metrics}\label{sec_3b}
Six evaluation metrics are considered in this study, where three (i.e., Dice, HD and ASSD) require ground-truth masks to be provided and measure the region-wise similarity, and three (i.e., NCC, MI and MSE) are unsupervised and measure the image-wise similarity. The region-wise similarity is more direct than image-wise similarity, since the goal of image registration is to find the spatial correspondence between moving and fixed image. We only calculate the image-wise similarity of the two images in the foreground area (voxel intensity is greater than zero) of the fixed image. Note that values of Dice, HD and ASSD are calculated for each RoI individually and the averaged values of all RoIs are reported. 

For the registration involves two modalities, we first calculate image-wise similarity for each modality and then average them. Taking $MSE$ as an example, it can be formulated as:

\begin{equation}
	\begin{aligned}
	\frac{1}{2}(&MSE(I_m^{T1w}\circ \Phi_{aff} \circ\Phi_{def}, I_f^{T1w}) + \\
	&MSE(I_m^{T2w}\circ \Phi_{aff} \circ\Phi_{def}, I_f^{T2w}) ).
	\end{aligned}
\end{equation}
This process is the same for the other two image-wise similarity metrics, i.e., NCC and MI. The calculation details of the six metrics are listed below:

\subsubsection{Dice score}
The Dice score of two regions is defined in Eq. \eqref{eq_8}. The range of Dice score is from 0 to 1. When two images are registered perfectly, the Dice score is 1.

\subsubsection{Hausdorff Distance (HD)}
The Hausdorff Distance measures the maximum mismatch between surfaces of two volumes, which is formulated as:
\begin{equation}
	H D(A, B)=\max _{a \in A}\left\{\min _{b \in B}\{||a - b||\}\right\},
\end{equation}
where A and B indicate all voxels on the surfaces of two volumes respectively.

\begin{table*}\centering
	\caption{Comparison results on IXI dataset(weakly supervised). The best performance is marked in bold.}
	\label{table1}
	\resizebox{\textwidth}{!}{
		\begin{tabular}{|l|c|c|c|c|c|c|c|} 
			\hline
			Method& Dice & HD & ASSD& NCC& MI& MSE& Para(M)\\
			\hline
			Elastix\cite{cite10}& $0.580\pm0.055$& $2.631\pm0.550$& $0.565\pm0.131$& $0.9141\pm0.0116$& $0.5530\pm0.0084$& $19.806\pm5.206$ & $-$\\
			ANTs\cite{cite9,cite42}& $0.657\pm0.040$& $1.550\pm0.170$& $0.378\pm0.067$& $0.9207\pm0.0104$& $0.5425\pm0.0061$& $13.957\pm1.447$& $-$\\
			VoxelMorph\cite{cite15}& $0.734\pm0.039$& $1.084\pm0.147$& $0.246\pm0.062$&$0.9341\pm0.0075$ &$0.5571\pm0.0061$ & $13.375\pm1.411$& $\bm{14.47}$\\
			VoxelMorph-diff\cite{cite16}& $0.732\pm0.031$& $1.111\pm0.122$& $0.260\pm0.048$&$0.9421\pm0.0065$ &$0.5566\pm0.0062$ & $13.166\pm1.421$& $14.50$\\
			Dual-PRNet\cite{cite22}& $0.714\pm0.059$& $1.153\pm 0.363$& $0.285\pm0.130$ &$0.9267\pm0.0071$ &$0.5568\pm0.0063$ & $13.912\pm1.390$& $15.03$\\
			VTN\cite{cite21}& $0.706\pm0.061$& $1.154\pm0.307$& $0.288\pm0.117$&$0.9120\pm0.0091$ &$0.5500\pm0.0065$ & $14.943\pm1.230$& $98.92$\\
			2x10 cascade VTN\cite{cite30}& $0.750\pm0.023$& $1.052\pm0.095$& $0.224\pm0.041$&$0.9439\pm0.0064$ &$0.5578\pm0.0062$ & $13.103\pm1.384$& $296.69$\\
			Ours& $\bm{0.761\pm0.020}$& $\bm{1.035\pm0.069}$& $\bm{0.209\pm0.030}$&$\bm{0.9455\pm0.0069}$ &$\bm{0.5588\pm0.0062}$ & $\bm{12.877\pm1.543}$& $15.40$ \\
			\hline
	\end{tabular}}
	\label{tab_1}
\end{table*}

\subsubsection{Average Symmetric Surface Distance (ASSD)}
Similar to HD, ASSD measures the averaged mismatch between surfaces of two volumes, which is formulated as:

\begin{footnotesize}
\begin{equation}
		ASSD(A, B)=\frac{1}{|A|+|B|}\left(\sum_{a \in A} \min _{b \in B}||a-b||+\sum_{b \in B} \min _{a \in A}||b-a||\right),
\end{equation}
\end{footnotesize}
where A and B indicate all voxels on the surfaces of two volumes respectively.

\subsubsection{Normalized cross-correlation(NCC)}
It is a commonly used similarity metric in image registration. A higher NCC indicates the two images are well-aligned. The NCC of the two images is defined in Eq. (\ref{eq_6}).

\subsubsection{Mutual Information(MI)}
The MI value of the two images A, B is written as:
\begin{equation}\label{eq_12}
	MI(A, B)=H(A)+H(B)-H(A,B),
\end{equation}
where $H(A)$ and $H(B)$ represent the information entropy of image A and image B respectively, and $H(A, B)$ is the joint entropy of A and B. A higher MI indicates the two images are well-aligned.

\subsubsection{Mean Squared Error(MSE)}
The MSE value of the two images A, B is written as:
\begin{equation}
	MSE(A,B)=\frac{1}{N}\sum{|A-B|^2}.
\end{equation} 
A lower MSE indicates the two images are well-aligned.



\section{Experimental Results}

\subsection{State-of-the-arts methods and experimental details}\label{sec_4a}
We compare our method with two open-sourced traditional registration toolkits, i.e., Elastix\cite{cite10} and ANTs\cite{cite11}, and learning-based state-of-the-arts including VTN\cite{cite22}, 2x10 cascade VTN\cite{cite28}, VoxelMorph\cite{cite16}, VoxelMorph-diff\cite{cite17} and Dual-PRNet\cite{cite23}. For Elastix and ANTs, we use the same commands with those used in \cite{cite22}. VTN, 2×10-cascade VTN, VoxelMorph and VoxelMorph-diff have released their source codes. Dual-PRNet\footnote{\url{https://github.com/OldDriverJinx/reimplemention-of-Dual-PRNet}} is re-implemented by ourselves due to the unavailable source code. Note that, VTN and 2×10-cascade VTN can be treated as progressive registration only, Dual-PRNet can be treated as coarse-to-fine estimation only, and VoxelMorph and VoxelMorph-diff can be treated as direct estimation without decomposition of deformation field.

At the top of each comparison learning-based method, we cascade a same affine network as our implementation, which can be jointly trained in an end-to-end fashion. All learning-based methods are trained under the same training setting. Specifically, we utilize Adam optimizer for minimizing the loss, set the number of epochs to 10 with the learning rate initialized to 10$^{-4}$, and set the batch size to 1 image pair per batch. The learning rate is halved twice during the training after the 4$^{th}$ and the 7$^{th}$ epoch, respectively. 

\begin{figure}[t]
	\centerline{\includegraphics[width=1.0\columnwidth]{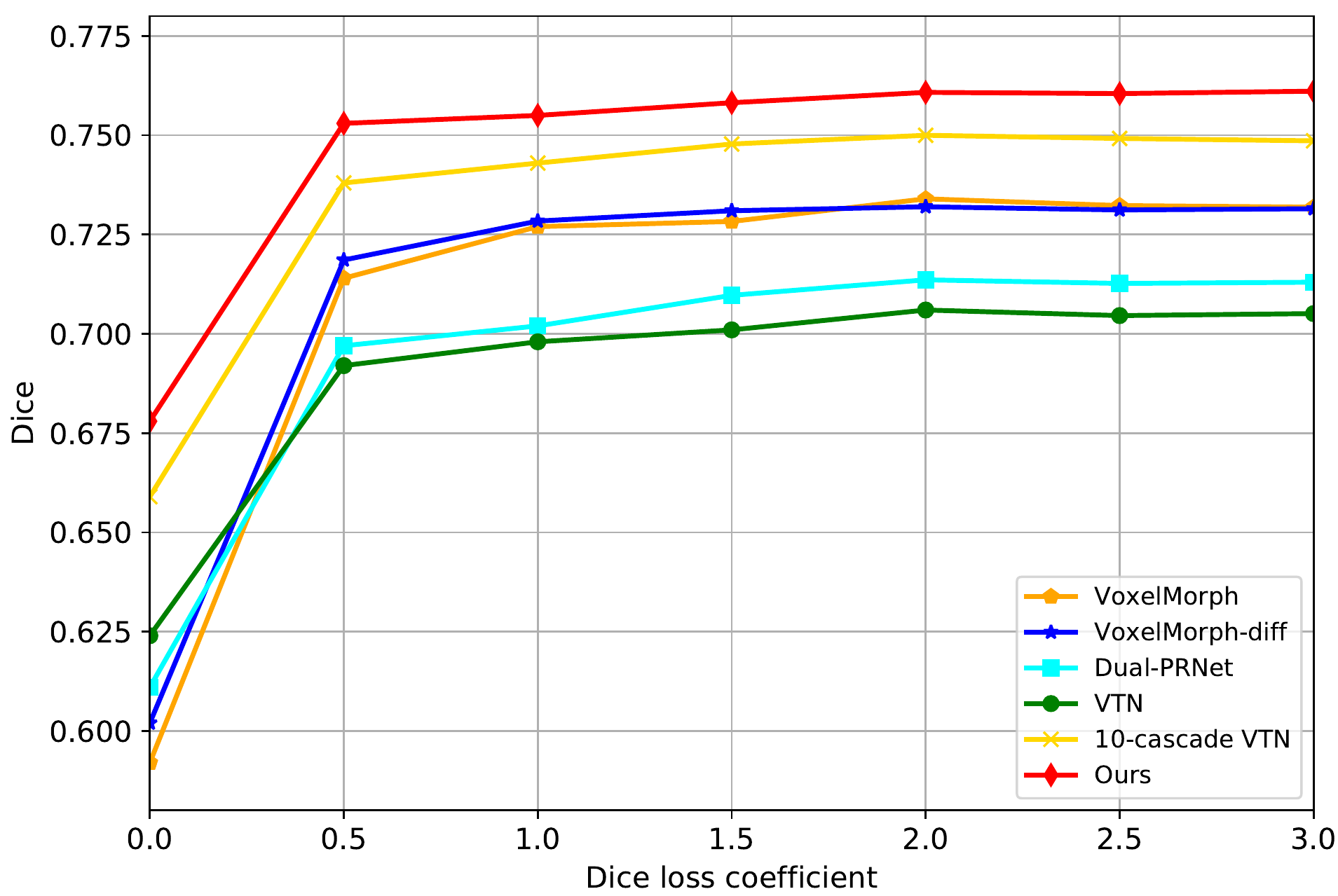}}
	\caption{Dice of different methods under different values of Dice loss $\alpha_{4}$} 
	\label{fig_7}
\end{figure}

For each dataset, the specific experimental details are presented as follows:

\begin{itemize}
	\item \textbf{IXI Dataset:} We randomly select 521 and 55 samples for training and testing respectively. The high-quality MRI data in CIT168 is utilized as the moving image. All methods are trained to register the concatenated T1w and T2w images of CIT168 to those in IXI. We train all methods in a weakly supervised manner by optimizing Eq. (\ref{eq_10}) with the same coefficient setting (i.e., $\alpha_{1}$, $\alpha_{2}$, $\alpha_{3}$, $\alpha_{4} =0.1,1,1,2$). The number of batches per epoch is 3,000. The total number of batches is thus 30,000. 

	\item \textbf{Private Dataset:} We randomly select 80 samples from our private dataset for fine-tuning all compared methods trained on the IXI dataset, and test them on the rest 20 samples. The number of batches per epoch is 1,000. The total number of batches is thus 10,000. We train all methods in a weakly supervised manner by optimizing Eq. (\ref{eq_10}), and the coefficient setting is consistent with that for IXI.
	\item \textbf{LPBA Dataset:} We utilize a collection of datasets including ADNI (66 T1w images) \cite{cite39}, ABIDE (1,287 T1w images) \cite{cite40}, ADHD (949 T1w images) \cite{cite41} to train all compared methods using the consistent coefficient setting with that for IXI and our private dataset but excluding $L_{seg}$ due to unavailable segmentations in the collection of datasets. To be consistent with the experimental settings in 2×10-cascade VTN \cite{cite30}, we train the methods to register each collected data (i.e., the moving image) to the first image in LPBA (i.e., the fixed image), and each batch of the collected data is augmented by a random B-Spline field of 5×5×5 control points within a maximum displacement of 12. The number of batches per epoch is 10,000. The total number of batches is thus 100,000. In the phase of testing, the first image in LPBA is treated as the fixed image as in the phase of training, while the rest 39 images in LPBA are treated as moving images to warp. All 56 labeled anatomical structures on LPBA are used to evaluate the region-wise similarity, i.e., Dice, HD and ASSD.
\end{itemize}

\begin{figure}[!t]
	\centerline{\includegraphics[width=1.0\columnwidth]{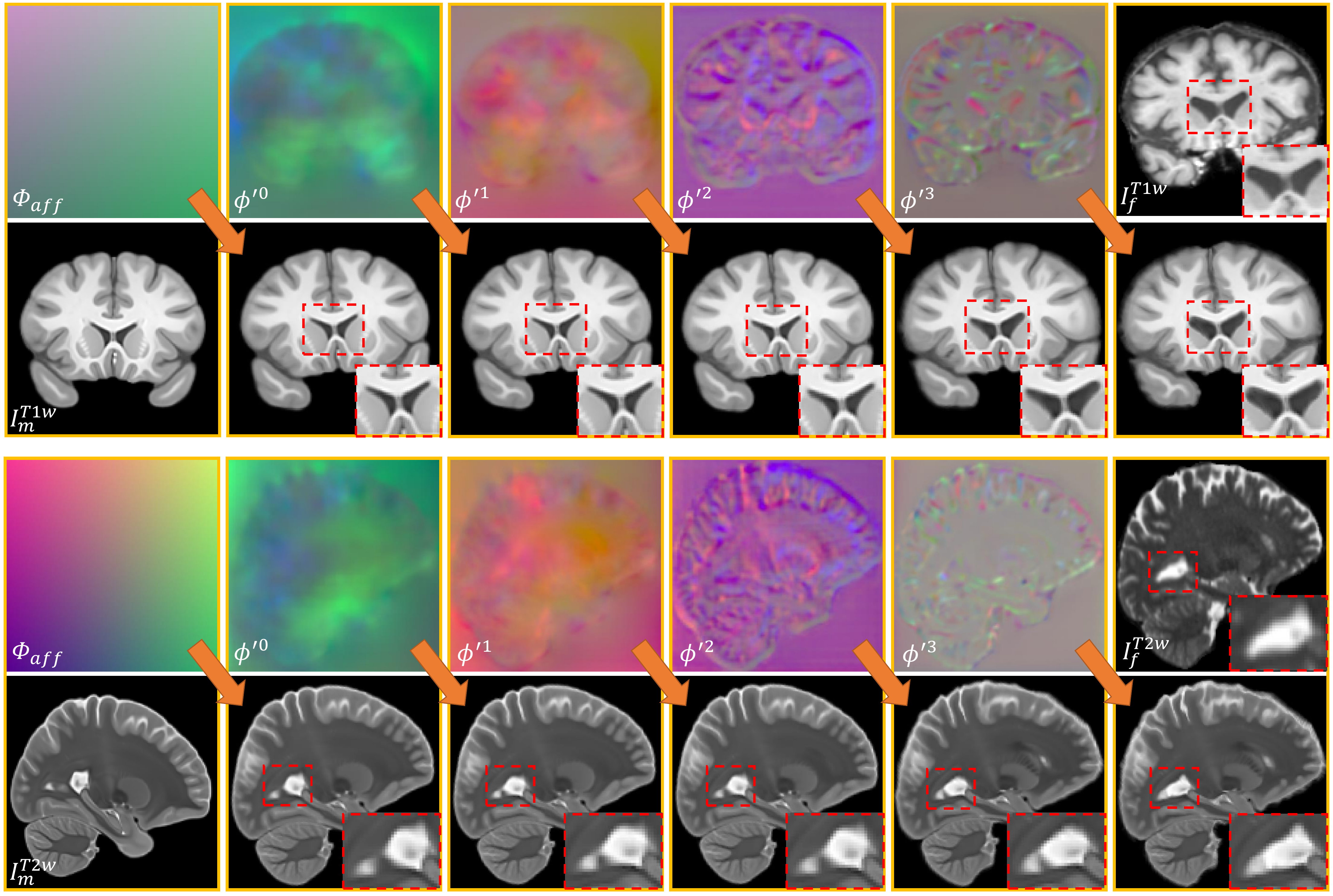}}
	\caption{Our progressive and coarse-to-fine deformation process from different views and different modality (coronal T1w in the first two rows and sagittal T2w in the last two rows). The sub-field on the top deforms the  image (only show slice for illustration) on the bottom, yielding the moved one on the bottom-right. The fixed image is presented at the top-rightmost as the reference.}
	\label{fig_8}
\end{figure}

\begin{table*}\centering
	\caption{Comparison results on our private dataset(weakly supervised). The best performance is marked in bold.}
	\label{table2}
	\resizebox{\textwidth}{!}{
		\begin{tabular}{|l|c|c|c|c|c|c|} 
			\hline
			Method  & Dice & HD & ASSD & NCC & MI & MSE \\
			\hline
			Elastix\cite{cite10}& $0.620\pm0.028$& $2.050\pm0.366$& $0.465\pm0.065$& $0.9329\pm0.0053$& $0.5551\pm0.0047$& $11.810\pm0.963$\\
			ANTs\cite{cite9,cite42}& $0.718\pm0.014$& $1.441\pm0.104$& $0.295\pm0.020$& $0.9358\pm0.0060$& $0.5548\pm0.0043$& $11.386\pm0.862$\\
			VoxelMorph \cite{cite15} & $0.738\pm0.028$& $1.124\pm0.119$ & $0.258\pm0.049$&$0.9424\pm0.0056$ & $0.5642\pm 0.0039$& $10.475\pm0.793$\\
			VoxelMorph-diff\cite{cite16}& $0.739\pm0.025$& $1.126\pm0.107$& $0.260\pm0.038$&$0.9472\pm0.0050$ &$0.5639\pm0.0038$ & $10.363\pm0.763$\\
			Dual-PRNet\cite{cite22} & $0.718\pm0.037$ & $1.167\pm 0.149$ & $0.288\pm0.060$& $0.9362\pm0.0053$ & $0.5638\pm0.0038$& $11.345\pm0.802$\\
			VTN\cite{cite21}&$0.706\pm0.042$ &$1.191\pm0.142$ &$0.304\pm0.071$& $0.9219\pm0.0067$ &$0.5573\pm0.0042$ & $12.608\pm0.908$ \\
			2x10-cascade VTN\cite{cite30}&$0.753\pm0.023$ &$1.080\pm0.067$ &$0.231\pm0.028$& $0.9538\pm0.0045$ &$0.5658\pm0.0040$ & $9.897\pm0.795$ \\
			Ours & $\bm{0.762\pm0.020}$ &$\bm{1.046\pm0.077}$ & $\bm{0.214\pm0.025}$&$\bm{0.9543\pm0.0046}$ & $\bm{0.5658\pm0.0040}$ & \bm{$9.720\pm0.757}$\\
			\hline
			
	\end{tabular}}
	\label{tab_2}
\end{table*}

\begin{figure*}[!t]
	\centerline{\includegraphics[width=0.9\textwidth]{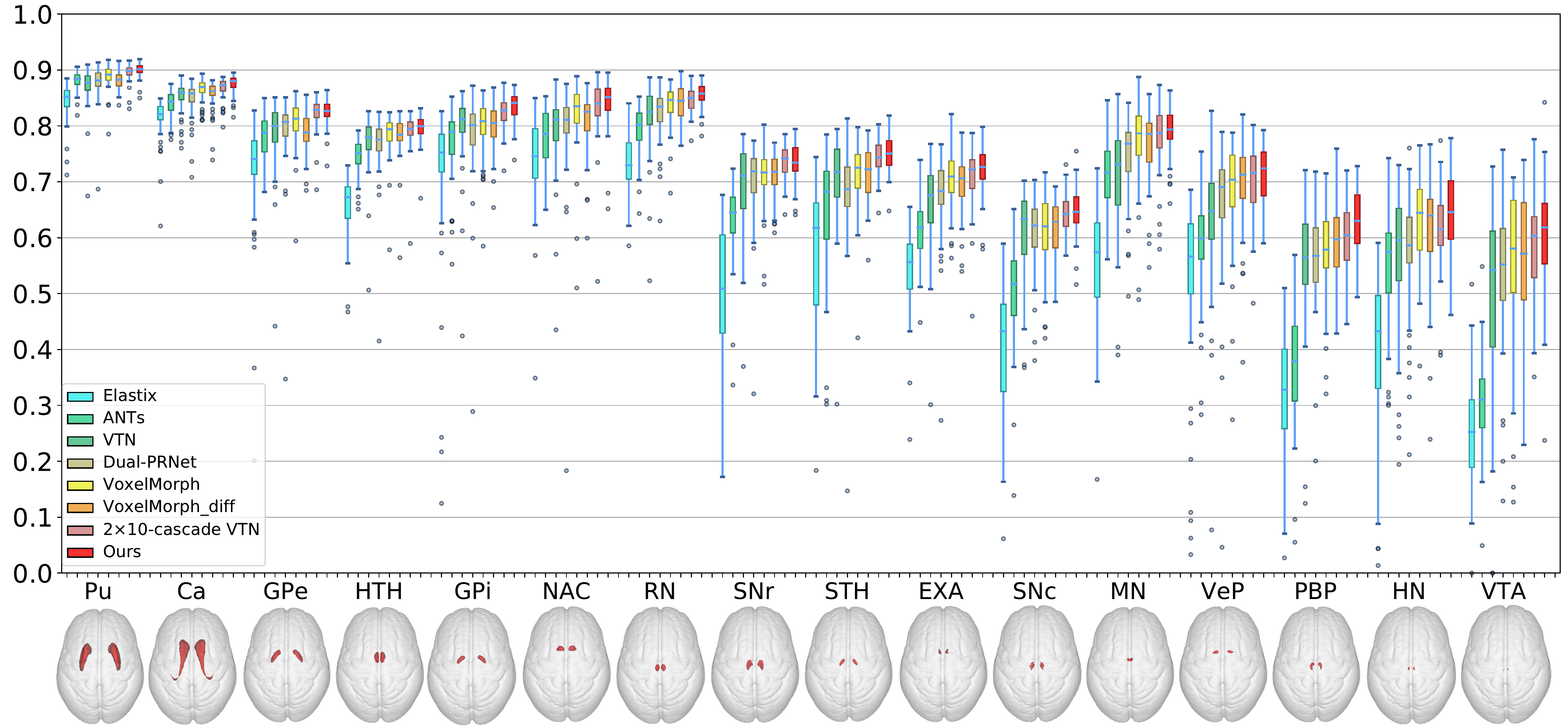}}
	\caption{Boxplots of Dice scores of 16 subcortical nuclei for comparison of methods. The subcortical nuclei phantoms are presented under the names and are ranked by the average numbers of the region voxels in decreasing order.}
	\label{fig_9}
\end{figure*}


\subsection{Comparison with the state-of-the-arts on IXI}\label{sec_4b}

We first heuristically tune $\alpha_{4}$ in Eq. (\ref{eq_10}) to verify the effect of Dice loss on the performance. The other three coefficients in Eq. (\ref{eq_10}) are fixed since they are inherited from \cite{cite21,cite30}. We adjust $\alpha_{4}$ from 0 to 3 with an equal interval of 0.5. Fig. \ref{fig_7} illustrates the registration performance of Dual-PRNet, VTN, 2×10-cascade VTN, VoxelMorph, VoxelMorph-diff as well as our method in terms of Dice under different $\alpha_{4}$. From Fig. \ref{fig_7}, we can observe that (1)	our method consistently surpasses other five methods regardless of the settings of $\alpha_{4}$ , and (2) as the value of $\alpha_{4}$ increases, the performance of all methods first increases, and then reaches the optimal with $\alpha_{4}$ equal to 2. Therefore, we set the hyperparameters to $\alpha_{1}, \alpha_{2}, \alpha_{3},\alpha_{4} = 0.1, 1, 1, 2$ in the following experiments.

We show some visualization samples in Fig. \ref{fig_8} to illustrate the progressive registration process by our predicted coarse-to-fine deformation sub-fields \{$\phi'^{0},\cdots, \phi'^{3}$\}, which are converted from \{$v'^{0},\cdots, v'^{3}$\} by the scheme defined in Eq. (\ref{eq_2}). For a better visualization, each sub-field is colorized by treating x-, y-, z-direction map as red-, green-, blue-channel respectively. As can be seen, each sub-field is more detailed than the previous one. With the predicted deformation field changes from coarse to fine, the moving image deforms progressively to be more similar to the fixed image.

Tab. \ref{tab_1} gives the quantitative comparison results with other state-of-the-art methods. A higher NCC, MI, Dice, and a lower MSE, HD, ASSD indicates a more accurate registration result. From those comparison results, we can have three observations:

(1) By comparing Dice, HD and ASSD values in the  $2^{nd}$ to $4^{th}$ columns, we can find that our method guarantees the most consistent RoIs, i.e., 16 subcortical nuclei, between the fixed and moved images comparing to other methods. This represents that our method can better guide the neurosurgeon to trace brain RoIs by reforming itself into an atlas-based segmentation method. Moreover, comparing with VTN and Dual-PRNet, which are progressive only and coarse-to-fine only, our proposed joint decomposition is more effective and improves the average Dice coefficient by around 8\% at most.

(2) By comparing NCC, MI and MSE values in the 5$^{th}$ to 7$^{th}$ columns, it is demonstrated that the registered images by our method have the highest image-wise similarity than those obtained by other methods. This implies that most regions with the same anatomic semantics are well-aligned in our results.

(3) The last column gives the memory occupation of parameters of different methods. Our method sacrifices a little bit of computational cost comparing to Dual-PRNet and VoxelMorph (less than 1 Megabyte), but achieves the best registration performance. The usage of VTN and 2x10-cascade VTN is 7 times and 20 times larger than the others' respectively, and it is mainly due to their multi-stage strategy for decomposition, where each stage is an isolated CNN model.

\begin{table*}\centering
	\caption{Comparison results on lpba dataset(unsupervised). The best performance is marked in bold, and the secondary best performance is marked by a symbol ‘*’.}
	\resizebox{\textwidth}{!}{
		\begin{tabular}{|l|c|c|c|c|c|c|c|} 
			\hline
			Method & Dice & HD & ASSD & NCC & MI & MSE& Para(M) \\
			\hline
			Elastix\cite{cite10} &$0.674\pm0.013$& $4.953\pm0.326$ & $1.258\pm0.069$&$0.9749\pm0.0028$ & $\bm{0.6150\pm0.0002}$ &$9.564\pm0.659$& $-$ \\
			ANTs\cite{cite9,cite42} &$0.708\pm0.015$& $4.885\pm0.359$ & $1.138\pm0.082$&$0.9889\pm0.0030$ & $0.6135\pm0.0003$ &$6.441\pm0.894$& $-$ \\ 
			VoxelMorph\cite{cite15}&$0.694\pm0.014$ &$4.872\pm0.312$ &$1.200\pm0.078$&$0.9878\pm0.0015$ &$0.6131\pm0.0003$ &$6.936\pm0.585$ & $\bm{14.47}$ \\
			VoxelMorph-diff\cite{cite16} &$0.692\pm0.015$ &$4.921\pm0.321$ &$1.212\pm0.080$&$0.9878\pm0.0015$ &$0.6132\pm0.0003$ &$6.941\pm0.567$& $14.50*$ \\
			Dual-PRNet\cite{cite22} &$0.709\pm0.013$& $4.692\pm0.333$ & $1.111\pm0.075$&$0.9901\pm0.0028$ & $0.6136\pm0.0002$ &$6.164\pm0.797$& $15.03$ \\
			VTN\cite{cite21} &$0.703\pm0.014$ &$4.969\pm0.333$ & $1.286\pm0.076$&$0.9877\pm0.0013$ &$0.6129\pm0.0002$ &$6.869\pm0.489$&$98.92$ \\
			2×10-cascade VTN\cite{cite30} &$0.719\pm0.012$ &$4.713\pm0.333$ & $1.069\pm0.067$&$0.9942\pm0.0004$ &$0.6128\pm0.0002$ &$4.962\pm0.337$&$296.69$\\
			\hline
			Ours &$0.721\pm0.014*$ & $4.656\pm0.318*$ &$1.059\pm0.071*$& $0.9945\pm0.0028*$ &$0.6138\pm0.0002*$ &$4.612\pm0.954*$&$15.40$ \\
			2-cascade Ours &$\bm{0.722\pm0.014}$ & $\bm{4.653\pm0.339}$ &$\bm{1.045\pm0.073}$& $\bm{0.9960\pm0.0025}$ &$0.6137\pm0.0002$ &$\bm{3.959\pm0.958}$& $16.62$ \\
			\hline
	\end{tabular}}
	\label{tab_3}
\end{table*}

\begin{figure*}[!t]
	\centerline{\includegraphics[width=1.8\columnwidth]{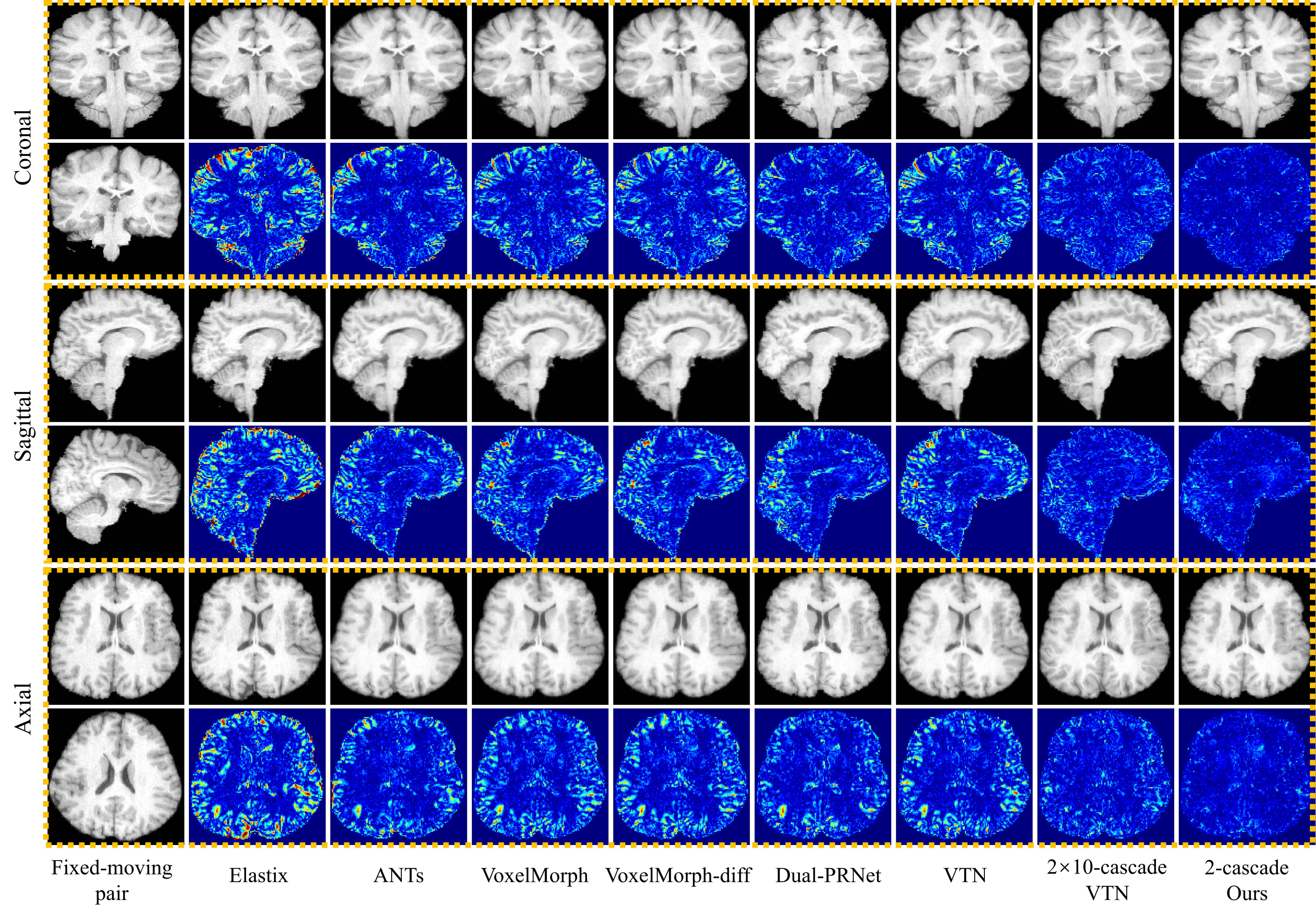}}
	\caption{Visualization results of different registration methods in three views. In each view, the two images of the first column are moving image and fixed image respectively, and the other columns are deformed moving image with its corresponding subtraction image of different methods.} 
	\label{fig_10}
\end{figure*}

Based on above three observations, it is verified that our method can improve both image-wise similarity and region-wise similarity, with a comparable requirement of computational cost to the coarse-to-fine method and more lightweight than the progressive method. 

Fig. \ref{fig_9} shows the boxplots of the Dice scores in 16 subcortical nuclei regions obtained by different methods. As can be seen, a superior performance of registration is consistently achieved by our method even at the two smallest regions. Moreover, the Dice scores obtained by our method are more stable (short length of the box) while others fluctuate as the number of region voxels decreases, verifying the robustness of our method.

\subsection{Comparison with the state-of-the-arts on the private dataset}\label{sec_4c}
We examine the adaptability of comparison methods in this subsection by fine-tuning the models trained on the IXI dataset using the training data of our private dataset, and then testing the performance of each method. The comparison results are shown in 2$^{nd} $ to 8$^{th}$ rows of Tab. \ref{tab_2}. As can be seen, our method produces registration results with the highest image-wise similarity measured by NCC, MI and MSE, and the best consistency of RoIs evaluated by Dice, HD and ASSD, which demonstrates a better adaptability of our method to newly collected datasets. Surprisingly, VoxelMorph is the third best among all comparison methods while it just treats the registration task as a segmentation task and employs a vanilla segmentation CNN (i.e., U-Net) without designing any interpretable module. This raises a question that \emph{will the supervision of those RoIs contribute a lot to the high performance of segmentation-style VoxelMorph.} We answer it in the next subsection.

\subsection{Comparison with the state-of-the-arts on LPBA}\label{sec_4d}

To answer the previous question, we examine the registration performance without supervision of ground-truth RoI masks, that is, all methods are trained in an unsupervised fashion. All evaluation metrics are reported in Tab. \ref{tab_3}. From those results, we find that the performance of two versions VoxelMorph degrade and are the worst two among the compared learning-based methods, implying that direct usage of segmentation model could be a good ``medicine" for registration but also acts as a ``poison" if the supervision of RoIs is unavailable. This conclusion is also evidenced in Fig. \ref{fig_7}, where VoxelMorph ranks last when the Dice loss is disabled (i.e., $\alpha_4=0$), and improves significantly with $\alpha_4$ greater than 0.

\begin{table}\centering
	\caption{Smoothness comparison of different methods. Areas with negative Jacobian determinant are considered folding. The fraction of folding is presented in E-notation (e.g., 1e-2=0.01).}
	\resizebox{1.0\columnwidth}{!}{
		\begin{tabular}{|l|c|c|c|} 
			\hline
			Method& Jacobian Std. & Folding	$\left(\%\right)$&Dice \\
			\hline
			Elastix\cite{cite10}&$0.189\pm0.022$ & $0.00\pm0.00$& $0.674\pm0.013$ \\
			ANTs\cite{cite9,cite42}&$0.193\pm0.012$ & $0.00\pm0.00$& $0.708\pm0.015$ \\
			Dual-PRNet\cite{cite22}&$0.906\pm0.059$ & $1.61e$-$1\pm2.44e$-$2$&$0.709\pm0.013$  \\
			VoxelMorph\cite{cite15}&$0.247\pm0.057$ &$5.15e$-$3\pm6.80e$-$3$&$0.694\pm0.014$  \\
			VoxelMorph-diff\cite{cite16}&$0.191\pm0.016$ &$0.00\pm0.00$&$0.692\pm0.015$ \\
			VTN\cite{cite21}&$0.179\pm0.024$ &$0.00\pm0.00$&$0.703\pm0.014$ \\
			2×10-cascade VTN\cite{cite30}&$0.355\pm0.068$ &$1.22e$-$6\pm7.54e$-$6$&$0.719\pm0.012$  \\
			Ours& $0.289\pm0.008$ &$3.58e$-$4\pm1.03e$-$4$&$0.721\pm0.014$ \\
			\hline
	\end{tabular}}
	\label{tab_4}
\end{table}

As can be seen from Tab. \ref{tab_3}, our method achieves the best performance compared with other state-of-the-arts in terms of all metrics except MI. It is worth noting that by comparing the 8$^{th}$ and the 9$^{th}$ rows of Tab. \ref{tab_3}, the Dice value of our method is comparable to that of the sophisticated 2×10-cascade VTN but with nearly 20 times fewer parameters, which well demonstrates the effectiveness of our proposed joint progressive and coarse-to-fine registration method.

We also cascade our deformable registration network twice (2-cascade Ours) following the inference strategy utilized by \cite{cite30}. The numbers of parameters of the affine and deformable parts in our method are 14.18 M and 1.22 M, respectively. Therefore, cascading our model only slightly increases the number of parameters from 15.40 M to 16.62 M. As can be seen from the last two rows of Tab. \ref{tab_3}, cascading our method further improves the registration performance in terms of all metrics except MI, and especially decreases MSE by 14.2\%.


Fig. \ref{fig_10} shows visual examples of the comparison results. In contrast, our deformed images are closest to the fixed one, yielding more clean subtraction images. The subtraction image is obtained by calculating the absolute value of difference between the warped and the fixed images for every voxel. The red color in the subtraction image indicates large difference of voxel intensity, and the blue color means the difference is small.

In medical image registration, it is very important to ensure the smoothness of the deformation field. We evaluate the smoothness of the predicted deformation fields by two metrics, i.e., the \emph{standard deviation of Jacobian determinants} at all voxels in each image and the \emph{fraction of folding} which is the percentage of folding voxels whose Jacobian determinants are negative. The transformation defined by the deformation field is diffeomorphic if and only if the percentage of folding voxels is zero \cite{cite54}. Therefore, lower standard deviation of the Jacobian determinant and percentage of folding voxels indicate a better smoothness of predicted deformation field. Tab. \ref{tab_4} presents the smoothness metrics of our method compared to other baseline methods. 

As can be seen in Tab. \ref{tab_4}, among these methods, Elastix, ANTs, VTN and VoxelMorph-diff guarantee the diffeomorphic property (with no folding voxels) but significantly sacrifices the Dice value of those brain regions. Comparing with 2×10 cascade VTN whose Dice value is comparable to ours, our predicted deformation fields have a lower standard deviation of Jacobian determinants and a competitive folding percentage (around 7.5 folding voxels in a 128×128×128 deformation field of ours vs. 2.6 of 2×10 cascade VTN). These results demonstrate that our method can achieve a superior registration performance with a promising smoothness of the predicted deformation field, i.e., a diffeomorphic transformation for more than 99.99\% voxels of the predicted deformation field.

Besides the diffeomorphic property of the deformation field, we also verify the invertibility of the affine transformation matrix by calculating its determinant. We found that those determinants of predicted matrixs on all datasets scatter around 1, and none of them is equal or even close to 0. That is, the predicted matrix by the affine registration network is invertible for every testing sample.

\begin{figure}[!t]
	\centerline{\includegraphics[width=0.8\columnwidth]{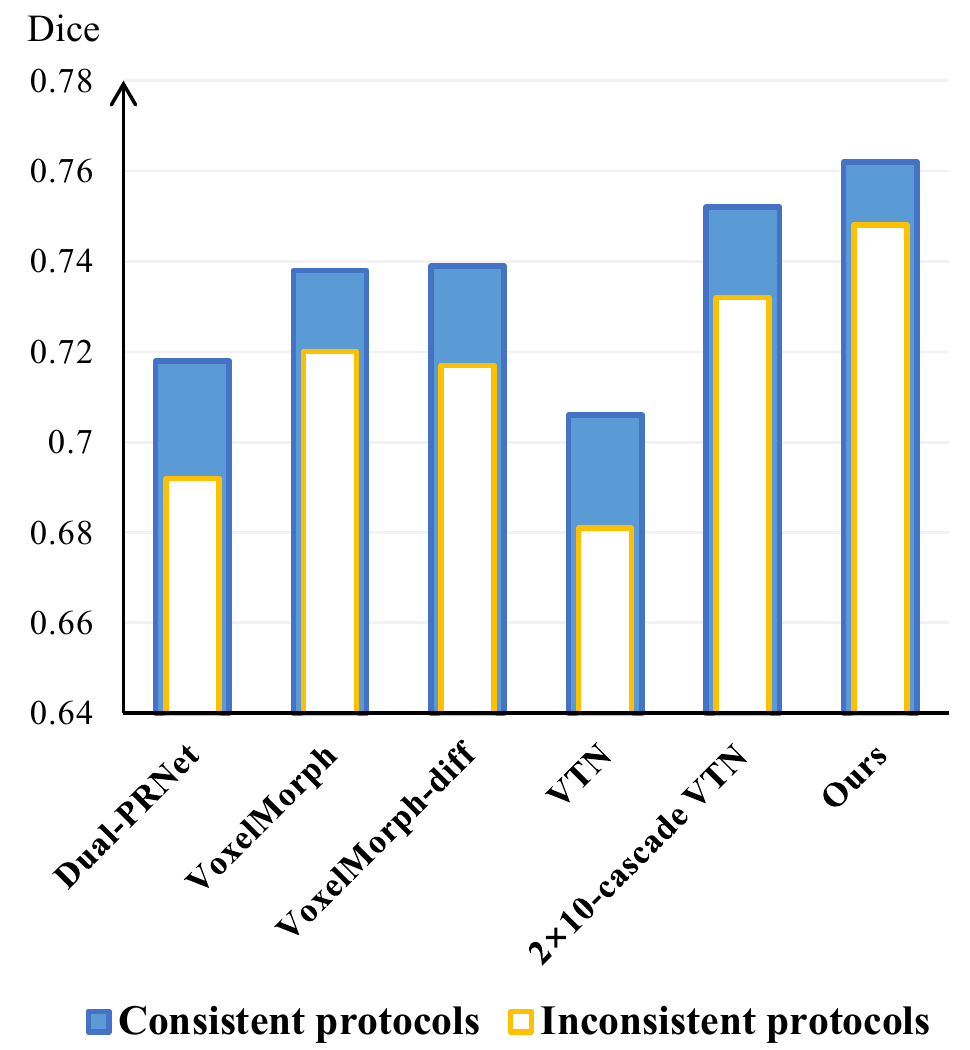}}
	\caption{Generalization test of the SOTA learning-based registration methods on private dataset.}
	\label{fig_11}
\end{figure}

\subsection{Generalization Experiment}\label{sec_4e}

The performance of a deep learning model might decrease due to the inconsistent distribution of the testing data and training data. In this section, we discuss the generalization of comparison methods under two situations, i.e., (1) \emph{inconsistent scanning protocols}, and (2) \emph{inconsistent fixed images}, between phases of training and testing.

\begin{table*}\centering
	\caption{Results of our four variant methods on IXI dataset. '\checkmark' and '-' indicate enabling and disabling respectively. The best performance is marked in bold.}
	\resizebox{\textwidth}{!}{
		
		\begin{tabular}{|p{20pt}<{\centering} |p{20pt}<{\centering}|c|c|c|c|c|c|} 
			\hline
			DFI & NFF  & Dice & HD & ASSD& NCC& MI  & MSE\\
			\hline
			-& - 		  			& $0.721\pm0.055$& $1.113\pm0.261$& $0.263\pm0.103$& $0.9352\pm0.0071$& $0.5561\pm0.0064$& $13.762\pm1.471$\\
			\checkmark& - 			& $0.735\pm0.028$& $1.053\pm0.101$& $0.237\pm0.042$& $0.9421\pm0.0063$& $0.5576\pm0.0062$& $13.321\pm1.518$\\
			-& \checkmark 			& $0.739\pm0.031$& $1.068\pm0.150$& $0.235\pm0.050$& $0.9404\pm0.0065$& $0.5571\pm0.0064$& $13.436\pm1.547$\\
			\checkmark& \checkmark  & $\bm{0.761\pm0.020}$& $\bm{1.035\pm0.069}$& $\bm{0.209\pm0.030}$& $\bm{0.9455\pm0.0069}$ &$\bm{0.5588\pm0.0062}$ & $\bm{12.877\pm1.543}$\\
			\hline
	\end{tabular}}
	\label{tab_5}
\end{table*}

For the situation of inconsistent scanning protocols, we compared the performance of different learning-based methods with and without fine-tuning, and the evaluation results are illustrated in Fig. \ref{fig_11}. The white bar represents the Dice coefficient achieved by using models trained on IXI to directly perform registration on our private dataset without fine-tuning (i.e., inconsistent protocols), and the blue bar represent the Dice coefficient achieved by using the training images of our private dataset to fine-tune the model (i.e., consistent protocols). Therefore, the gap between the blue and white bar represents the performance drop when scanning protocol in the test phase is inconsistent with that in the training phase. A smaller gap between the white and blue bars indicates a better generalization performance.

Comparing with the results after fine-tuning, Dice achieved by our method only drops by 0.014, while other methods degrade, with Dice dropping by 0.018, 0.022, 0.025, 0.020, and 0.026 for VoxelMorph, VoxelMorph-diff, VTN, 2×10-cascade VTN, and Dual-PRNet, respectively. These results demonstrate a compromising generalization achieved by our method, for the different scanning protocols between phases of training and testing.


\begin{figure}[!t]
	\centerline{\includegraphics[width=0.8\columnwidth]{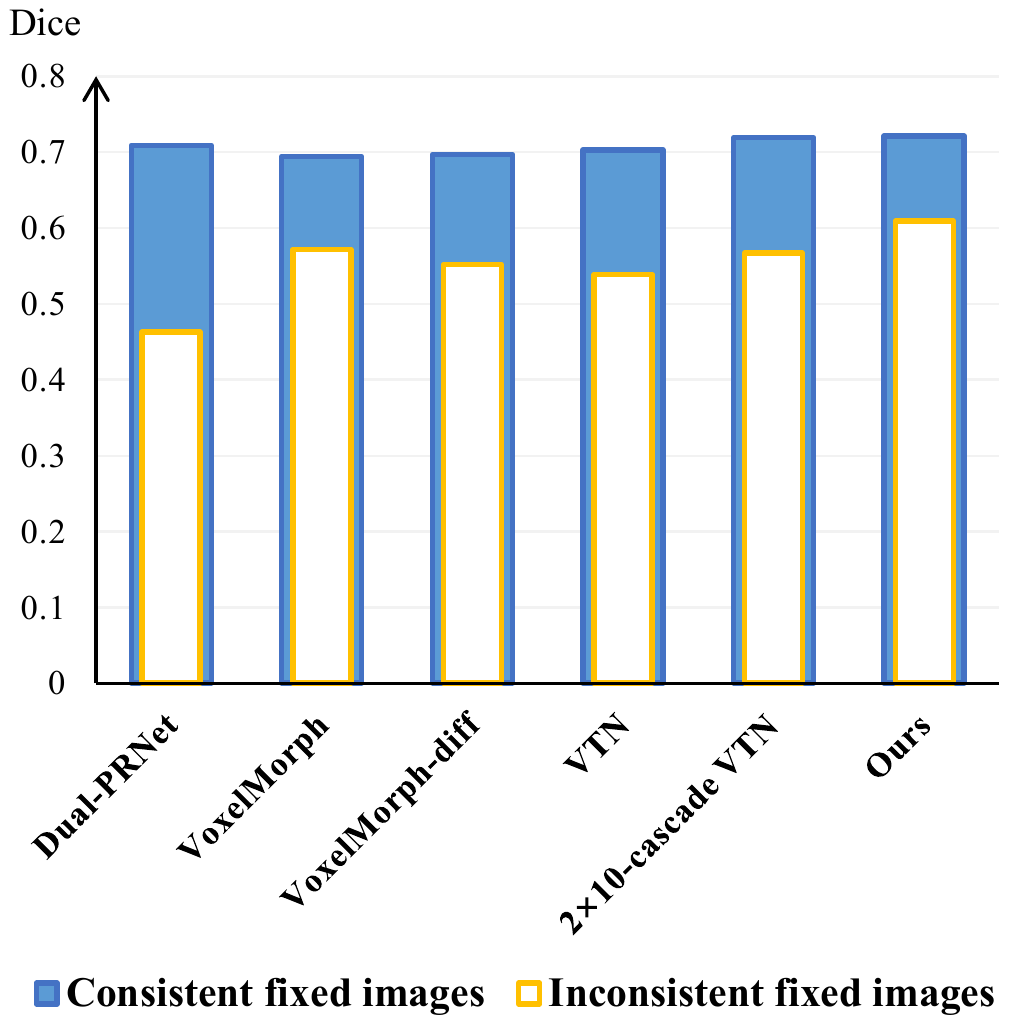}}
	\caption{Generalization test of the SOTA learning-based registration methods on LPBA dataset.}
	\label{fig_12}
\end{figure}

For the second situation of inconsistent fixed images, we first trained all learning-based methods including ours to align images (i.e., moving image) in ADNI, ABIDE and ADHD with the first image (i.e., fixed image) in LPBA, that is, the fixed image is always the same one. In the phase of testing, we directly used the trained model to register all the 40 images in LPBA in a pairwise manner instead of using the same fixed image for training. The pairwise registration means every two images are selected from LPBA, and treated as fixed and moving images, respectively. Thus, 40 images of LPBA totally construct 40×39=1560 fixed-moving image pairs to register. 

The registration performance using consistent fixed images (i.e., 39 image pairs to register) and inconsistent fixed images (i.e., 1560 image pairs to register) is evaluated for all methods, and the evaluation results are illustrated in Fig. \ref{fig_12}. 

In comparison, the Dice coefficient of our method decreases by 0.112, while it drops by 0.122, 0.145, 0.164, 0.152, and 0.246 for VoxelMorph, VoxelMorph-diff, VTN and 2×10-cascade VTN, and Dual-PRNet, respectively. These results demonstrate a compromising generalization of our method when the fixed image differs from that in the training phase and keeps changing.

From both Fig. \ref{fig_11} and Fig. \ref{fig_12}, we can observe that the performance of Dual-PRNet drops drastically with Dice decreasing by 0.026 in the first situation and by 0.246 in the second situation, even though it also utilizes a dual-encoder similar with ours. We believe the main reason is that Dual-PRNet not only separates the encoding process, but also splits the decoding process. That is, the processes of feature extraction on the fixed-moving image pair are independent in Dual-PRNet from the beginning to the end, which negatively affects the generalization ability for the features are extracted completely independently.

Interestingly, the superior generalization results of our method under the two situations demonstrate that dual-encoder plus single-decoder could be a better alternative comparing with dual-encoder plus dual-decoder (i.e., Dual-PRNet) and single-encoder plus single-decoder (i.e., VoxelMorph, VTN, etc.). These results are consistent with our high-level motivation for using dual-encoder plus single-decoder, that is, we would like to learn more alignment-friendly encoding features for the prediction of velocity fields using dual-encoder, and to extract more shared decoding features from both fixed and moving image using single-decoder, preventing the model from overfitting to a specific intensity pattern. Therefore, these two together can lead to the competitive registration performance and the promising generalization ability.

\subsection{Ablation Studies}\label{sec_4f}

In this subsection, we aim at evaluating the effectiveness of our proposed two key modules, i.e., DFI and NFF. We develop four versions of our method which are trained with disabling DFI and/or disabling NFF. Specifically, with disabling DFI, those sub-fields are integrated as they are without being reweighted. With disabling NFF, feature maps from three sources are fused as they are without channel-wise and spatial-wise attentions. The ablation study is conducted based on the IXI dataset and the T1w and T2w images of CIT168 are utilized as the atlas (moving image). Tab. \ref{tab_5} shows the registration results of the four versions of our method.

By comparing the first row of Tab. \ref{tab_5} and the results in Tab. \ref{tab_1}, we can find that even without reweighting and attention, our method achieves a superior performance over VTN and Dual-PRNet, which again demonstrates the effectiveness of joint decomposition. DFI with reweighting boosts the Dice by 1.9\% (see the 1$^{st}$ and 2$^{nd}$ rows) and NFF with attention improves the Dice by 2.5\% (see the 1$^{st}$ and 3$^{rd}$ rows), which implies that a proper feature fusion plays an important role. Combining both modules, the Dice further increases by 3.6\% and 3.1\% receptively, verifying that DFI and NFF are not mutually excluded and can positively affect each other during the learning.

We have also performed the statistical test on the comparison between the baseline variant (i.e., the second row in Tab. \ref{tab_6}) and each of the three variants (i.e., the last three rows in Tab. \ref{tab_6}). The method of statistical test we used is paired-samples T test which is implemented in the SPSS software.  Tab. \ref{tab_6} gives the p-values of all comparisons in terms of all evaluation metrics. We can conclude from the comparison results shown in the 2$^{nd}$ to 4$^{th}$ row of Tab. \ref{tab_5} and the corresponding p-values shown in the 2$^{nd}$ to 3$^{rd}$ row of Tab. \ref{tab_6}, that adding either DFI or NFF can improve the registration accuracy significantly (p-value$<$ 0.05). Moreover, we find that most p-values in the last row of Tab. \ref{tab_6} are dramatically smaller than others in the 2$^{nd}$ to 3$^{rd}$ rows, which demonstrates that DFI and NFF can cooperate to further improve the registration accuracy significantly.

\begin{table}\centering
	\caption{The statistical test of the comparison between the three variants (w/ DFI, w/ NFF or w/ both) and the baseline variant (w/o both). The p-values are presented in E-notation (e.g., 1e-2=0.01).}
	\resizebox{1.0\columnwidth}{!}{
		
		\begin{tabular}{|c|c|c|c|c|c|c|c|} 
			\hline
			DFI & NFF  & Dice & HD & ASSD& NCC& MI  & MSE\\
			\hline
			\checkmark& - 		   & $3.22e$-$3$& $3.86e$-$3$& $3.86e$-$3$& $2.71e$-$25$& $1.20e$-$19$& $7.16e$-$25$\\
			-& \checkmark 		   & $1.19e$-$5$& $7.53e$-$3$& $3.82e$-$4$& $2.65e$-$29$& $1.73e$-$24$& $1.67e$-$3$\\
			\checkmark& \checkmark & $1.70e$-$9$& $3.77e$-$3$& $2.57e$-$6$& $3.50e$-$34$ &$1.30e$-$26$ & $6.10e$-$25$\\
			\hline
	\end{tabular}}
	\label{tab_6}
\end{table}

\section{Conclusion}
In this paper, we propose a unified framework for decomposing a target deformation field into a series of multi-scale deformation sub-fields via two novel modules, i.e., DFI and NFF. DFI dynamically integrates all previous coarser sub-fields to a single field, and NFF utilizes the integrated field to progressively align both moving and fixed encoding feature maps, and fuses the aligned feature maps and the last decoding feature map with channel and spatial attentions to help estimate a finer sub-field afterwards. DFI and NFF are used alternately and repeatedly in decoding blocks, factorizing the full-size deformation field into sub-field in both progressive and coarse-to-fine manners jointly. Comparing with progressive registration only, our method enjoys more light-weight decoding blocks instead of CNN models in separate stages, and thus makes the optimization more efficient. Comparing with the coarse-to-fine estimation only, our method splits the difficult task to those each of which a decoding block can handle rather than counting on every two adjacent layers, and thus makes the decomposition more effective. We conduct extensive and comprehensive experiments including four brain MRI datasets and seven state-of-the-art registration methods. Comparison results on IXI demonstrate a superior performance  of our method over other state-of-the-arts in terms of both image-wise similarity and region-wise similarity. Experimental results on our private dataset verify the adaptability of our method. Comparison results on LPBA show the effectiveness of our joint decomposition comparing with progressive only, coarse-to-fine only and direct estimation. The generalization test verifies our employed dual-encoder plus single-decoder could be a better alternative for preserving the registration performance when scanning protocols and input image patterns change between training and testing. The ablation study also well demonstrates the significant performance gains brought by our proposed novel DFI and NFF modules. Our future work will include the extension to more modalities with huge style discrepancy, e.g., T1w and DWI, and more organs with intricate structures, e.g., small intestine.

\section*{Acknowledgment}
This work was supported in part by National Natural Science Foundation of China (Grant No. 62006087), Fundamental Research Funds for the Central Universities (2021XXJS033), Science Fund for Creative Research Group of China (Grant No. 61721092), Director Fund of WNLO, Research grants from United Imaging Healthcare Inc.

\bibliographystyle{IEEEtran}
\bibliography{inference}

\end{document}